\documentclass[11pt,letterpaper]{article}
\usepackage{jheppub}
\usepackage{comment,xcolor}
\usepackage{amsmath}
\usepackage{mathtools}
\usepackage{amssymb}
\usepackage{amsfonts}
\usepackage{amsthm}
\usepackage{mathrsfs}
\usepackage[all]{xy}
\usepackage{tensor}
\usepackage{footnote}
\usepackage{graphicx} 
\usepackage[justification=raggedright,singlelinecheck=false]{caption}
\usepackage{subcaption}
\usepackage{cleveref}
\usepackage{booktabs}

\newtheorem{thm}[equation]{Theorem}
\newtheorem{lem}[equation]{Lemma}
\newtheorem{conj}[equation]{Conjecture}

\newtheorem{defn}[equation]{Definition}

\newtheorem{rem}[equation]{Remark}

\newcommand{\dd}{\mathrm{d}}

\newcommand{\beq}{\begin{equation}}
\newcommand{\eeq}{\end{equation}}

\numberwithin{equation}{section}

\DeclareMathOperator*{\argmin}{argmin}

\def\R{\mathcal{R}}
\def\A{\mathcal{A}}
\def\B{\mathcal{B}}

\def\O{\overline{\Omega}}
\def\M{\mathcal{M}}

\def\l{\tilde{\ell}}
\def\S{\mathcal{S}}
\def\PMS{\mathcal{P}}
\newcommand*{\tr}{\mathrm{tr}}
\def\D{\mathcal{D}}

\title{Outer entropy = Bartnik-Bray inner mass, and the gravitational ant conjecture}

\author{Jinzhao Wang }
\affiliation{\small \it Institute for Theoretical Physics, ETH 8093 Z\"urich, Switzerland}
\emailAdd{jinzwang@phys.ethz.ch}

\abstract{
Entropy and energy are found to be closely tied on our quest for quantum gravity. We point out an interesting connection between the recently proposed outer entropy, a coarse-grained entropy defined for a compact spacetime domain motivated by the holographic duality, and the Bartnik-Bray quasilocal mass long known in the mathematics community. In both scenarios, one seeks an optimal spacetime fill-in of a given closed, connected, spacelike, codimension-two boundary. We show that for an outer-minimizing mean-convex surface, the Bartnik-Bray inner mass matches exactly with the irreducible mass corresponding to the outer entropy. The equivalence implies that the area laws derived from the outer entropy are mathematically equivalent as the monotonicity property of the quasilocal mass. It also gives rise to new bounds between entropy and the gravitational energy, which naturally gives the gravitational counterpart to Wall's ant conjecture. We also observe that the equality can be achieved in a conformal flow of metrics, which is structurally similar to the Ceyhan-Faulkner proof of the ant conjecture. We compute the small sphere limit of the outer entropy and it is proportional to the bulk stress tensor as one would expect for a quasilocal mass. Lastly, we discuss some implications of taking quantum matter into consideration in the semiclassical setting.}

\begin{document}
\maketitle

\section{Introduction}

The outer entropy is initially proposed by Engelhardt and Wall (EW)~\cite{engelhardt2018decoding,engelhardt2019coarse} as a coarse-grained entropy for black hole. Motivated by the Jaynes' principle of maximum entropy~\cite{jaynes1957information,jaynes1957information2}, a coarse-grained entropy for black hole could be defined as the maximal entropy over what we do not know inside the horizon while holding fixed what we can observe in the exterior. Formally, the outer entropy of an apparent horizon $\Sigma$ in asymptotically AdS spacetime is given by
\begin{equation}
\S(\Sigma) := \sup_\rho S_{\mathrm{vN}}(\rho) : D(\O)\,\, \text{fixed}\label{formalouter}
\end{equation}
where $S_{\mathrm{vN}}(\rho) $ denotes the von Neumann entropy of the boundary quantum $\rho$ dual to the classical geometry characterized by some initial data set $(N,h,K),$ where $N=\Omega\cup_\Sigma\O$ and $h,K$ are the first and second fundamental forms. $D(\O)$ is the domain of dependence of $(\O,h,K)$ on a partial Cauchy slice $\O$ connecting $\Sigma$ to the boundary $\B$ ($\partial\O=\Sigma\cup \B$), and we call it the \emph{outer wedge} $O_W(\Sigma)$. Let us also define the interior spacetime that we would like to maximize the entropy over as the Cauchy development of the other half of the initial data $(\Omega,h,K),$ which we refer as the \emph{fill-in} and its Cauchy development as the {\it inner wedge} $I_W(\Sigma):=D(\Omega)$. The EW construction is essentially motivated by the the holographic duality. In particular, the Hubeny-Ryu-Rangamani-Takayanagi (HRRT) prescription~\cite{ryu2006holographic,ryu2006aspects,hubeny2007covariant,rangamani2017holographic} implies that one can treat the coarse-grained entropy over quantum states on the boundary as the optimization over geometric data in the bulk. We will later review the precise bulk definition. EW proved that the outer entropy of an apparent horizon is given by the $\text{Area}(\Sigma)/4G_N\hbar,$ consistent with what we know about the black hole thermal entropy. Besides the matching value, this variational formulation of the entropy has a statistical interpretation that immediately yields the area laws of the spacelike and null holographic screens~\cite{bousso2015new,bousso2015proof}. Soon after EW's work, Bousso, Nomura and Remmen (BNR)~\cite{nomura2018area,bousso2019outer} generalized the EW method to solve for the outer entropy of untrapped surfaces, which we henceforth dub as the \emph{BNR algorithm}.

In general relativity, there are other instances of such optimization-over-geometries construction that are not motivated by considerations of entropy or the holographic principle. In search of a good definition quasilocal mass associated with a codimension-2 surface $\Sigma$, there are several proposals that involving optimizations. One example is a promising proposal by Wang and Yau~\cite{wang2009quasilocal}, where they seek to minimize the Hamiltonian over permissible isometric embeddings of the target surface into the Minkowski reference. Here, we focus on another quasilocal mass definition that has been proposed and studied for a few decades, which resembles in many aspects with the outer entropy. This is called the Bartnik-Bray mass~\cite{bartnik1989new,bartnik1997energy,bray2001proof,bray2004penrose,szabados2009quasi}. Motivated by the ADM mass and the positive mass theorem~\cite{schoen1979proof,schoen1981proof}, Bartnik proposed~\cite{bartnik1989new} a definition of quasilocal mass, the \emph{Bartnik mass} $M_B(\Sigma)$, via looking for a minimal-ADM-mass extension of a given compact spacetime domain $(\Omega,h,K)$ bounded by $\Sigma=\partial \Omega$. Intuitively, one can think of the Bartnik mass as the ``quasilocalized'' ADM mass.

Our main object of interest is the dual version of the Bartnik mass given by Bray~\cite{bray2001proof,bray2004penrose}. In 1999, based on his proof of the \emph{Riemannian Penrose Inequality} (RPI)~\cite{bray2001proof}, Bray proposed a slightly modified version of the Bartnik mass, named as the \emph{Bartnik-Bray outer mass}, $M_{\text{outer}}(\Sigma)$, together with its dual, called the \emph{Bartnik-Bray inner mass}, $M_{\text{inner}}(\Sigma),$ where the constraint and optimization domain are exchanged and the infimum is replaced by the supremum. The inner mass is defined as the irreducible mass corresponding to the maximal area of the minimal surface homologous to the given $\Sigma.$ Note that Bray used {\it inner} referring to the variational region, whereas EW used {\it outer} referring to the fixed region. Albeit the opposite names, the authors actually proposed the same optimization construction up to different conditions and motivations. Physically speaking, both variational definitions try to fill the interior of a surface with the largest black hole. They can be categorized as instances of the \emph{fill-in problem}, which has been studied by geometers in different contexts~\cite{bray2004penrose,jauregui2013fill,jauregui2013extensions,mantoulidis2018capacity}. Without ambiguity, we will now simply refer them as the \emph{inner mass} and the \emph{outer mass}. We shall point out that BNR also realize the outer entropy can be a good candidate for quasilocal mass due to its monotonicity~\cite{bousso2019outer}, and the AdS version is proposed in \cite{engelhardt2019holographic}. However, the outer entropy by definition can also depend on the data on the outer wedge, so it is not a quasilocal quantity~\cite{szabados2009quasi}. In this work, we show that this insight can be traced back to Bray and Bartnik~\cite{bartnik1989new,bray2001proof}, and the outer entropy can be quasilocalized for an \emph{outer-minimizing} surface.

We shall start by reviewing the definitions of the outer entropy and Bartnik-Bray quasilocal mass in section \ref{sec:outer} and \ref{sec:qlm}. In section~\ref{sec:equi} we show that for an outer-minimizing mean-convex surface, the outer entropy and the inner mass are equivalent. Fundamental insights can be drawn from this connection and we will discuss them in section~\ref{sec:implication}. For example, the area laws associated with the holographic screens are mathematically equivalent to the monotonicity of the Bartnik-Bray quasilocal mass. The equivalence also leads to several interesting inequalities relating the outer entropy and the purely gravitational energy. In particular, we show that given some cut $\Sigma$ on an initial data set $(N,h,K),$ the Penrose inequality implies the infimum of the total mass of the spacetime over all possible extensions is lower bounded by the outer entropy $\S(\Sigma)$. We move on to conjecture that the equality holds, which in many ways resembles the ant conjecture due to Wall concerning matter fields~\cite{wall2017lower}. We also show that the equality can be asymptotically approached under a conformal flow of metrics, in a way that is structurally similar to Ceyhan and Faulkner's proof of the ant conjecture, via the Connes cocycle flow~\cite{ceyhan2018recovering}. We believe this correspondence is not merely a coincidence. Our conjecture provides a new perspective on the question of why one can only depict gravitational energy quasilocally rather than locally, (unlike the energy of ordinary matter, which is represented by a local energy-momentum tensor).  Furthermore, under simplifying assumptions, we compute the small sphere limit of the outer entropy and inner mass using the BNR algorithm in section~\ref{sec:ssl}. We found that the small sphere limit of the outer entropy is given by the local bulk stress tensor, exactly as one would expect for a quasilocal mass. Lastly in section~\ref{sec:quantum}, we consider the possiblity of elevating the classical gravitational ant conjecture to semiclassical regime by adding the contribution of energy and entropy of matter. We in turn propose a quantum Penrose inequality that is worth further considerations. We finish with some discussions on future directions in section~\ref{sec:discuss}. 

Let us state some assumptions and fix the notations before we proceed. We work in the classical limit in the bulk, which could be the large N strong coupling limit of the AdS/CFT correspondence~\cite{maldacena1999large,witten1998anti,gubser1998gauge} or else some speculative form of flat space holography\footnote{The same generalisations are also used by BNR~\cite{nomura2018area,bousso2019outer}.}~\cite{bagchi2010correspondence,li2011holography,bagchi2012flat,bagchi2015entanglement,jiang2017entanglement,compere2017kerr,nomura2017toward,nomura2018spacetime}.  We do not consider the quantum corrections to the HRRT prescription. We will be considering a topological codimension-two sphere $\Sigma$ embedded in a $n$-dimensional spacetime $(\M,g)$. We will consider both asymptotically flat (AF) and asymptotically hyperbolic (AH) initial data satisfying the dominant energy condition (DEC), as usually assumed in the studies of the Bartnik mass~\cite{anderson2019recent} to guarantee a well-posed initial value formulation\footnote{Instead of DEC, the weaker null energy condition (NEC) is assumed in earlier works on the outer entropy. Here we are treating things more generally so we stick to the DEC to avoid any subtle situations where the quantities we consider are ill-defined. In any case, our main result only explictly needs NEC.}.  $M(N,h,K)$ denotes the total mass of the data set. $A(\Sigma)$ denotes the area of $\Sigma\subset N$. We denote the outer and inner wedge of $\Sigma$ as $O_W(\Sigma), I_W(\Sigma)$ respectively. When there are multiple connected boundary components in the spacetime, we shall consider $\Sigma$ to enclose one chosen end with the conformal boundary $\B$. We allow the fill-ins $(\Omega,h,K)$ of $\Sigma$ to have multiple disconnected boundary components (ends). The metric on $\Sigma$ is $\gamma.$ The mean curvature of $\Sigma$ in $(N,h,K)$ associated with the outward-pointing unit normal $\boldsymbol{\nu}$ is denoted as $H$, and the mean curvature along $\Sigma$ of $N$ in $(\M,g)$ associated with the future-pointing unit normal $\mathbf{n}$ is denoted as $\tr_\Sigma K.$ We define $\mathbf{H}:=\tr_\Sigma K \mathbf{n} - H\boldsymbol{\nu}$ as the mean curvature vector of $\Sigma$ in $(\M,g)$. In the case of $N$ being null, $\mathbf{n},\mathbf{\nu}$ are replaced by the ingoing and outgoing null vectors $\ell^\pm$  normal to $\Sigma$,  and the associated mean curvatures are the expansion rates $\theta^\pm$. We say the surface is locally extremal if $\mathbf{H}=\mathbf{0}$, i.e. the null expansions vanish, $\theta^\pm=0$. The marginally outer-trapped surface (MOTS) is given by $\theta^+=0,\theta^-\leq 0.$ The apparent horizon on a Cauchy slice is the outermost MOTS on it.  We shall also use $\pm$ to indicate quantities associated with the ingoing/outgoing null congruences. 

We finish this section by defining the notions of \emph{outer-minimizing} and \emph{normal}.
\begin{defn}\label{outermini}
A topological codimension-two sphere $\Sigma$ is called outer-minimizing in $(N,h,K)$ if for any surface $\Sigma' \subset N$ enclosing $\Sigma$,
\begin{equation}
A(\Sigma)\leq A(\Sigma').
\end{equation}

Furthermore, a $\Sigma$ homologous to $\B$ is called outer-minimizing if there exists a partial Cauchy data $(\O,h,K)$ connecting $\Sigma$ to $\B$,  such that $\Sigma$ is outer-minimizing in $(\O,h,K)$.
\end{defn}

We shall address the significance of the outer-minimizing condition later. As an example, note that an apparent horizon is outer-minimizing in particular~\cite{bray2011pde,eichmair2010existence}.

\begin{defn}\label{normalsurface}
A topological codimension-two sphere $\Sigma$ is mean-convex if the mean curvature vector $\mathbf{H}$ is inward spacelike.
\end{defn}
The condition of $\mathbf{H}$ being inward spacelike means $H\geq 0$ and $|H|\geq |\tr_{\Sigma} K|$ with respect to any choice of embedding slice. Equivalently, one can demand the null expansions $\theta^\pm|_\Sigma$ to take definite but opposite signs on $\Sigma$, i.e. $\pm\theta^\pm\geq 0.$ Hence, in terms of the relativistic terminologies, the condition translates to $\Sigma$ being either untrapped or marginally trapped, or is termed as \emph{normal} surfaces in~\cite{bousso2019outer}.\footnote{In~\cite{bousso2019outer}, normal surfaces only refer to untrapped surfaces. Here, we extend the definition of ``normal'' to include the limiting cases.}. We will focus on outer-minimizing mean-convex surfaces and provide some contexts of these two conditions in section~\ref{sec:equi}.

\section{The outer entropy as a bulk quantity}\label{sec:outer}

The outer entropy is formally defined as (\ref{formalouter}). This is, however, not how the outer entropy is classically defined in the bulk, and the entropy is not evaluated directly through density operators in the previous works. Nevertheless, it is clear from the motivation above what the bulk definition should be. In the large N strong coupling limit, according to the HRRT prescription in holographic duality~\cite{ryu2006holographic,ryu2006aspects,hubeny2007covariant,rangamani2017holographic}, the entropy of the marginal quantum state dual to the classical bulk geometry is measured by the area of the locally extremal surface with the minimal area, called the Hubeny-Rangamani-Takayanagi (HRT) surface~\cite{hubeny2007covariant}. HRT surface can be identified via Wall's maximin prescription~\cite{wall2014maximin}.

\begin{defn}
Given a boundary subregion $\A,$ the HRT surface  $X(\A)$ of a boundary causal domain $D(\A)$ is defined as the minimal area surface on the maximal Cauchy slice\footnote{If there are multiple minimal area surfaces, any one can be the HRT surface. Here we only consider the maximizer $X(\A)$ that is stable. See~\cite{wall2014maximin} for more details.}
\beq
X(\A):= \sup_{N_{\A}} \argmin_{\substack{\sigma\subset N_{\A}\\ \sigma \in [\A]}} \,A(\sigma) \label{maximin}
\eeq
where the Cauchy slice $N_{\A}$ is anchored on $D(\A)$, $\sigma$ is anchored on $\partial\A$ and homologous to $\A$\footnote{$\sigma$ is not necessarily homologous to $\A$ via $N_{\A}$ as $N_{\A}$ could be anchored elsewhere in $D(\A).$ There is, however, a definition of \emph{restricted} maximin surface that requires $N_{\A}$ to be anchored at $\A$, which turns out to be equivalent as the original unrestricted version when maximin surface lies in a smooth region of spacetime~\cite{marolf2019restricted}.}.  ($[\A]$ denotes the homology class of $\A$.) The von Neumann entropy $S_\mathrm{vN}(\A)$ of the region $D(\A)$ is then measured by 
\beq
S_\mathrm{vN}(A)=\frac{A(X(\A))}{4G_N\hbar}.
\eeq
\end{defn}
In words, the HRT surface is located by first finding the minimal surface on a Cauchy slice and then maximizing over all Cauchy slices homologous to the boundary interval $\A.$ The existence conditions of the maximin have been given by Wall in~\cite{wall2014maximin} and extended in~\cite{marolf2019restricted}, so we can replace $\sup$ by $\max$ when considering horizonless spacetimes and black hole spacetimes with Kasner-like singularities following~\cite{wall2014maximin}. Here we keep it general because the equivalence that we seek to establish doesn't need the existence assumption.

In words, the outer entropy is measured by the area of the maximal HRT surface one can put into the inner wedge. When considering the outer entropy, we are taking a whole connected boundary component $D(\A)=\B$ as our causal domain. Hence, $\partial\A=\emptyset$ and the maximin surface is not anchored on the boundary, and $N_{\A}$ is just any  Cauchy slice so we shall remove the subscript. Let the outer wedge $O_W(\Sigma)$ be fixed, which is equivalent to fixing some partial Cauchy data $(\O,h_0,K_0),$ we have the following bulk definition for the outer entropy,

\begin{defn}\label{outerentropy}
The outer entropy of $\Sigma=\partial \O$ associated with the outer wedge data $(\O,h_0,K_0)$ is
\beq
\S(\Sigma)  := \sup_{(\Omega,h,K)} \frac{A(X(\B))}{4 G_N\hbar}=\sup_{(\Omega,h,K)}\max_{N\subset D(\Omega \cup_\Sigma \O)} \min_{\substack{\sigma\subset N\\ \sigma\in [B]}}\frac{A(\sigma)}{4 G_N\hbar}\label{outerformula}
\eeq
where $(\Omega,h,K)$ is the fill-in data that joins the fixed $(\O,h_0,K_0)$ at $\Sigma$ satisfying DEC and the following constraints:
\beq
\gamma |_{\Sigma_{in}} = \gamma |_{\Sigma_{out}};\,\,\,\theta^\pm |_{\Sigma_{in}} = \theta^\pm |_{\Sigma_{out}};\,\,\,\chi |_{\Sigma_{in}} = \chi |_{\Sigma_{out}}\label{junctionconditions}
\eeq
where $\chi := K(\cdot,\ell^-)$ is the twist or anholonomicity 1-form and $\ell^-$ is the ingoing null vector normal to $\Sigma$. 
\end{defn}
\begin{rem}
Here we only require DEC for the fill-in data without specifying anything about the matter sector as in EW~\cite{engelhardt2019coarse}. It would be interesting to fine-grain it depending on the relevant physical settings (cf. Discussion in BNR~\cite{bousso2019outer}).
\end{rem}

\begin{rem}
EW imposes (\ref{junctionconditions}) such that outer wedge and the inner wedge (fill-in) can be ``glued'' together properly~\cite{engelhardt2019coarse}. This is to ensure that the initial data on the entire Cauchy slice satisfies DEC in a distribution sense. 
\end{rem}

\begin{rem}\label{existence}
EW shows that for $\Sigma$ being the apparent horizon\footnote{\label{fn:minimar}EW generalizes the result on the apparent horizon~\cite{engelhardt2018decoding} to \emph{minimar} surfaces~\cite{engelhardt2019coarse}, which satisfy 1. $\Sigma$ is an outer-minimizing marginally trapped surface. 2. $\partial_+\theta^-<0.$ Both conditions ensure the extremal surface can be constructed via the method proposed by EW.}, the maximizer always exists and one can replace the $\sup$ with $\max$ in definition \ref{outerentropy}~\cite{engelhardt2018decoding}. Otherwise, for generic surfaces, we do not know if the maximizer exists.  
\end{rem}

The outer entropy is not quasilocal because it could depend on $O_W(\Sigma)$. In all the previous works~\cite{engelhardt2018decoding,engelhardt2019coarse,nomura2018area,bousso2019outer}, however, $\Sigma$ is demanded to be outer-minimizing (cf. Definition \ref{outermini}). This means any exterior surfaces on the initial data slice enclosing $\Sigma$ have area larger than $A(\Sigma).$ This condition, for example, is included in the minimar condition as required by EW~\cite{engelhardt2019coarse} (cf. foonote \ref{fn:minimar}). Then one can show that the HRT surface always lies within the inner wedge, $X\subset I_W(\Sigma)$ (cf. Lemma \ref{lem3}).  More importantly, an algorithm for evaluating the outer entropy is proposed by BNR for such surfaces. To our knowledge, the algorithm does not work for more general surfaces.

We shall say a few words about the BNR algorithm that computes the outer entropy. It uses the characteristic initial value formalism~\cite{rendall1990reduction,brady1996covariant,choquet2011cauchy,luk2012local,chrusciel2012many,chrusciel2014existence,chrusciel2015characteristic} to specify data on a null hypersurface $N_+$ fired towards the interior from $\Sigma.$ The data is chosen to have vanishing stress tensor and shear. EW proves~\cite{engelhardt2018decoding,engelhardt2019coarse} that a locally extremal surface exists on $N_+$ and it has the same area as the apparent horizon $\Sigma,$ which is the optimal case one can hope for. NR~\cite{nomura2018area} then generalizes the EW method to spherically symmetric outer-minimizing mean-convex surfaces. As a follow-up~\cite{bousso2019outer}, BNR tries to lift the spherical symmetry assumption, and then the locally extremal surface can be located subject to certain conditions. However, in this case the optimality of the chosen data is not proven but only argued. Hence, we do not know if the output of the algorithm with a generic input is the extremal surface with maximal area nor if the optimizer exists.

Now we can work directly with this classical bulk definition of the outer entropy, and we do not need the full holographic duality apparatus. We henceforth denote the HRT surface $X(\B)$ simply as $X$.

\section{The Bartnik-Bray quasilocal mass}\label{sec:qlm}

We first define the notion of spacetime extension~\cite{anderson2019recent}.
\begin{defn}\label{extension}
The permissible spacetime extension $\PMS_\Sigma$ of $\Sigma$ is the class of the asymptotically flat (hyperbolic) data $(\O,h,K)$  extending a given compact spacetime domain $(\Omega,h_0,K_0)$ with boundary $\Sigma=\partial \O=\partial\Omega,$ such that the complete manifold $\Omega \cup_\Sigma \O$ forms an initial data set that satisfies the dominant energy condition and $\Sigma$ is outer-minimizing.
\end{defn}
\begin{rem}\label{rem:non-empty}
It's difficult to determine if the Bartnik data admits an extension without constraining the data, and we don't know generally what are the necessary constraints on $h,K$ to make sure $\PMS_\Sigma$ is non-empty~\cite{anderson2019recent}. Here, we have a spacetime to start with and consider only Bartnik data induced on the chosen $\Sigma$, so $\PMS_\Sigma$ is non-empty by definition. 
\end{rem}
Bartnik argued~\cite{bartnik1997energy} that in order for the Hamiltonian and momentum constraints to be distributionally well-defined across $\Sigma,$ one should match
\beq\label{bbconditions}
\begin{aligned}
h |_{\partial \O} = \gamma &:= h_0 |_{\partial\Omega},\\
H_{\partial \O} = H &:= H_{\partial\Omega},\\
\tr_{\partial \O}K = k &:= \tr_{\partial\Omega}K_0,\\
\omega^\perp_{\partial \O}:=K(\cdot,\nu) = \omega &:= \omega^\perp_{\partial\Omega},
\end{aligned}
\eeq
where $h |_{\partial \O}$ is a Riemannian metric on $\Sigma$, $H_{\partial \O}$ is the mean curvature of $\Sigma$ in $\O$ with respect to the unit normal $\nu$, $\tr_{\partial \O}K $ is the mean curvature of $\O$ as embedded in the spacetime with respect to the unit normal $n$, $\omega^\perp_{\partial \O}$ is the connection one form, and similarly on the interior side induced by $(\Omega,h_0,K_0).$ This condition can also be obtained~\cite{bartnik1997energy} via demanding the boundary variation of the Regge-Teitelboim Hamiltonian~\cite{regge1974role} to vanish.

We define the tuple $(\Sigma,\gamma,H,k,\omega^\perp)$ as a \emph{Bartnik data} set. Given a domain $(\Omega,h_0,K_0)$, one can think of the Bartnik data being induced from it. However in general, the Bartnik data can be independently prescribed as a quasilocal data on $\Sigma$. The bartnik mass can be defined for either asymptotically flat extensions~\cite{bartnik1989new,bartnik1997energy,bray2001proof,bray2004penrose} or asymptotically hyperbolic extensions~\cite{pacheco2018asymptotically}.

\begin{defn}\label{outermass}
Given a Bartnik data set $(\Sigma,\gamma,H,k, \omega^\perp)$, the Bartnik-Bray outer mass is defined as~
\beq
M_{\text{outer}}(\Sigma,\gamma,H,k, \omega^\perp):=\inf_{(\O,h,K)\in \PMS_\Sigma} M(\O,h,K)\,.
\eeq
\end{defn}

\begin{rem}\label{rem:riemannian}
The Bartnik data and the outer mass also have a Riemannian version which is defined for the time-symmetric case $(K=0).$ Then the Bartnik data reduces to $(\Sigma,\gamma,H).$ 
\end{rem}

This definition is due to Bray~\cite{bray2001proof}. Bartnik originally proposed the definition for AF extensions, called the Bartnik mass $M_B$, that demands the extension to contain no horizons instead of imposing the outer-minimizing condition \cite{bartnik1989new,bartnik1997energy}. This slight variation in $\PMS_\Sigma$ is the only difference between the outer mass $M_{\text{outer}}$ and the Bartnik mass $M_B.$ Both of the original no-horizon condition  imposed by Bartnik and the outer-minimizing condition by Bray serve to rule out those ``bag of gold'' initial data sets. Otherwise, one can always hide $\Sigma$ behind some horizon such that the extension can have the ADM mass as small as possible~\cite{bartnik1989new,bray2004penrose,anderson2019recent}.

The Bartnik-Bray outer mass satisfies many desirable properties of quasilocal mass, such as positivity, rigidity and monotonicity, but it is very difficult to evaluate~\cite{szabados2009quasi} (cf.~\cite{anderson2019recent} for a survey of known results.).  Also, the spacetime Bartnik mass is much more tricky to analyze than its Riemannian counterpart. Based upon physical arguments, Bartnik conjectured that the minimizer always exists and is given by a stationary extension~\cite{bartnik1997energy}, but little is known about this conjecture~\cite{anderson2019recent}\footnote{\label{fn:BB}On the other hand, there are many results concerning the static extension conjecture, which is the Riemannian counterpart. It is recently proven to be false by Anderson and Jauregui~\cite{anderson2019embeddings}.}. When we restrict to horizons, the problem simplifies a bit. In the Riemannian setting, Mantoulidis and Schoen~\cite{mantoulidis2015bartnik} proved that the Bartnik mass $M_B$ (or $M_{\text{outer}}$) of a horizon is given by the irreducible mass (\ref{irr}). Their result is generalized to hyperbolic case (\ref{irrads}) by  Cabrera Pacheco, Cederbaum and Mccormick~\cite{pacheco2018asymptotically}. $M_{\text{outer}}$ is technically easier to work with than $M_B$. In the Riemannian setting, one can show that it is lower bounded by the Hawking mass~\cite{huisken2001inverse}, recovers the Schwarzschild mass for domains in the Schwarzschild spacetime containing the horizon and its small sphere limit can be evaluated~\cite{wiygul2018bartnik} (cf. Anderson~\cite{anderson2019recent}). Finally, we shall point out that  as compared to $M_{\text{outer}}$, the AdS version is much less studied in the literature. It was not studied only until recently been first proposed in~\cite{pacheco2018asymptotically}.

We now switch to the Bartnik-Bray inner mass $M_{\text{inner}}$. As we outlined above, the outer mass is essentially a problem concerning extensions, whereas the inner mass can be treated as a fill-in problem. To facilitate the definition of $M_{\text{inner}}$, we first need to define the fill-in of a Bartnik data set (see~\cite{jauregui2013fill} for the Riemannian version).
\begin{defn}
A fill-in of Bartnik data $(\Sigma,\gamma,H,k,\omega^\perp)$ is a compact, connected Riemannian codimension-one manifold $(\Omega,h,K)$ with boundary such that there exists isometric embedding $\imath: (\Sigma,\gamma,H,k,\omega^\perp)\hookrightarrow (\Omega,h,K)$ with the $\imath(\Sigma)$ being some connected component\footnote{Note that a fill-in $(\Omega,h,K)$ could have multiple ends, and we do need such fill-ins for non-trivial inner mass (see Remark \ref{rem:fillintype}).} of $\partial\Omega$ such that the induced $(h|_{\partial\Omega},H_{\partial\Omega},\tr_{\partial\Omega}K, \omega^\perp_{\partial\Omega})$ matches with the Bartnik data. We denote the set of admissible fill-ins as $\Gamma_\Sigma.$
\end{defn}
\begin{rem}
It's known that the mean curvature cannot be too large for $\Gamma_\Sigma$ to be non-empty~\cite{jauregui2013fill}. This is consistent with the fact that the BNR alogrithm doesn't output a fill-in if $-\theta^+\theta^-$, which is the norm of the mean curvature vector, is too large. Here, we have a spacetime to start with and consider only Bartnik data induced on the chosen $\Sigma$, so $\Gamma_\Sigma$ is non-empty by definition.
\end{rem}

A fill-in is thus an initial data set that can be legitimately inserted into the interior of $\Sigma$, whose domain of dependence is the inner wedge $I_W(\Sigma)$. It is commonplace to impose extra conditions on the Bartnik data set for the fill-in problem. For example, in the Riemannian case  (cf. Remark \ref{rem:riemannian}), the Bartnik data is given by a triple $(\Sigma,\gamma,H)$, and one usually demands $H$ to be a positive function~\cite{anderson2019recent}. Here we keep it general for the definition. Later we will impose mean-convexity (cf. Definition \ref{normalsurface}) to show the equivalence with the outer entropy.

Finally, we need to define the irreducible mass of a given area in asymptotically flat spacetime~\cite{christodoulou1971reversible,bray2009riemannian}
\beq
M_\mathrm{irr}(A) := \frac12 \left(\frac{A}{\Omega_{n-2}}\right)^{\frac{n-3}{n-2}}\label{irr}.
\eeq
 It sets a limit on the amount of energy that can be extracted from the black hole via the Penrose process. It can also be interpreted as the mass of a Schwarzschild black hole of the horizon area $A$. When $n=4$, we have the familiar expression $M_\mathrm{irr}(A)=\sqrt{\frac{A}{16\pi}}$. 
For the AdS case, we have a slightly different expression~\cite{bray2004penrose,mars2009present,itkin2012penrose}.
\beq
M_\mathrm{irr}(A) := \frac12 \left(\frac{A}{\Omega_{n-2}}\right)^{\frac{n-3}{n-2}}+\frac12\left(\frac{A}{\Omega_{n-2}}\right)^{\frac{n-1}{n-2}}\label{irrads}
\eeq
where we have set the AdS radius to 1.

The \emph{Penrose inequality} lower bounds the ADM mass $M$ of an initial data set by the irreducible mass. In arbitrary dimensions~\cite{bray2009riemannian}, it can be written as
\beq
M(N,h,K)\geq M_\mathrm{irr}(A(\Sigma_0))\label{PI}
\eeq  
for the apparent horizon (the outermost MOTS) $\Sigma_0$ on $(N,h,K),$ which is AF and satisfies DEC. The Penrose inequality also contains a rigidity statement: the equality holds if and only if $(N, h,K)$ is the initial value for the Schwarzschild spacetime. 

Similarly, in the AdS setting we have \emph{hyperbolic Penrose inequality} constraining an AH data if we replace the RHS on $\ref{PI}$ by $\ref{irrads}$. The LHS then denotes the total mass of an AH data~\cite{wang2001mass,chrusciel2003mass}. Also, the equality holds if and only if $(N,h,K)$ is the initial value for the AdS-Schwarzschild spacetime.

Though largely believed to be true, the general Penrose inequality is not proved yet. The RPI has been proved in dimensions less than eight~\cite{huisken2001inverse,bray2001proof,bray2009riemannian}. For the hyperbolic Penrose inequality, however, there is a holographic argument~\cite{engelhardt2019holographic}, which uses the outer entropy and an Euclidean path integral argument claiming the bulk dual to a maximum entropy state in a microcanonical
ensemble is the static AdS black hole~\cite{marolf2018microcanonical}. A mathematical proof of the hyperbolic Penrose inequality, even in the time-symmetric case, is still lacking. For more results and discussions on the Penrose inequality, see~\cite{mars2009present} for a comprehensive review.

The original definition of the inner mass is only given for four dimensional AF spacetime~\cite{bray2001proof,bray2004penrose}. Here we propose its natural $n$-dimensional version motivated by the formulation of the Penrose inequality in arbitrary dimensions~\cite{bray2009riemannian}, which also matches the proposal by BNR~\cite{bousso2019outer}.
\begin{defn}\label{inner}
Given a Bartnik data set $(\Sigma,\gamma,H,k, \omega^\perp)$, the Bartnik-Bray inner mass in an asymptotically flat spacetime is defined as
\begin{equation}
M_{\text{inner}}(\Sigma) :=\sup_{(\Omega,g,K)\in\Gamma_\Sigma}\min_{\substack{\sigma\subset \Omega,\\\sigma\in [\Sigma]}} M_\mathrm{irr}(A(\sigma))
\end{equation}
where the supremum is taken over all fill-ins $(\Omega,g,K)$ such that it satisfies the dominant energy condition and the minimum is taken over the homology class $[\Sigma]$.
\end{defn}

\begin{rem}\label{rem:fillintype}
Unlike the outer entropy \ref{outerentropy}, $M_{\text{inner}}$ is a quasilocal quantity as it only depends on the Bartnik data. Given some Bartnik data, there could be no valid fill-ins or only  compact fill-ins without boundary, and then we set $M_{\text{inner}} = 0.$ Hence, a non-degenerate inner mass is given by a fill-in that connects $\B$ to other ends, which we refer to as a non-trivial fill-in. The Riemannian case has been analyzed by Jauregui in~\cite{jauregui2013fill}, where it is proven that there is a threshold mean curvature value that determines if a non-trivial fill-in exists. 
\end{rem}
\begin{rem}\label{rem:existence}
Given a non-trivial fill-in (cf. Remark \ref{rem:fillintype}) of $\Sigma$ with inward-pointing mean curvature vector, the minimum in (\ref{inner}) can always be attained at some smooth minimal surface due to the results of Federer and Fleming~\cite{federer1960normal,federer2014geometric} (cf. Theorem 19 in~\cite{jauregui2013fill}). Regarding the supremum, we do not know the existence criteria on the Bartnik data~\cite{bartnik1997energy,anderson2019recent}. If the maximizer exists, then the minimal surface on it is locally extremal.
\end{rem}

The two versions of inner mass only differ in the definitions of irreducible mass, both of which are positive monotonically increasing functions of the area. Hence, they are essentially the same variational problem, which is to maximize the area. Apparently, the definition looks very similar to the maximin definition \ref{maximin} for the HRT surface of the whole boundary $\B$, up to the different homology classes. However, here the supremum is over all fill-ins, whereas for the HRT surface the Cauchy slices are constrained to evolve to the same spacetime $(\M,g).$ Hence, the same existence proof for the HRT surface does not work for $M_{\text{inner}}.$ Nevertheless, we will prove that it is equivalent to the outer entropy for an outer-minimizing mean-convex $\Sigma.$

\section{Equivalence}\label{sec:equi}
It is perhaps clear by now that the outer entropy and the inner mass look very similar to each other: they both search for the locally extremal surface with the maximal area, up to the different dimensions of the final quantities of interest. We call this area the \emph{supremum area}. In this section, we shall prove that the supremum areas for both optimization problems are identical, establishing the equivalence between the outer entropy and the inner mass:
\beq
M_\mathrm{inner}(\Sigma) = M_\mathrm{irr}(4\hbar G_N\S(\Sigma)).
\eeq

Let us first comment on the junction conditions required by EW (\ref{junctionconditions}). It is no surprise that these requirements are indeed given by the Bartnik data $(\Sigma,h,H,k,\omega^\perp)$ as well (\ref{bbconditions}). Note that under the null basis of the normal bundle $\{\ell^+,\ell^-\}$, the null expansions $(\theta^+, \theta^-)$ is the mean curvature vector and the twist is defined as $\chi_a := K(\cdot,\ell^-).$ Hence, the only difference is that (\ref{junctionconditions}) gives the continuity conditions specifically in terms of null frame variables, whereas the Bartnik data is given in a general form. They are the same up to a basis transformation. 
One can think of $(\Sigma,\gamma|_{\Sigma_{out}},\theta^\pm |_{\Sigma_{out}},\chi_a |_{\Sigma_{out}})$ induced from the outer wedge as a Bartnik data set given in the null frame.

Let us motivate why we demand $\Sigma$ to be both outer-minimizing and mean-convex in order to establish their equivalence. We first need to ``quasilocalize" the outer entropy. This equivalence argument is valid only because of the following lemma proved in~\cite{nomura2018area}, which we alluded to earlier. 
\begin{lem}\label{lem3}
For an outer-minimizing surface $\Sigma$, the HRT surface for the outer entropy, if it exists, always lies inside the inner wedge, $X\subset I_W(\Sigma)$.
\end{lem}
The outer-minimization thus quasilocalizes the outer entropy. Without the outer-minimizing condition, we might have the HRT surface being inside the outer wedge, then the two optimizers cannot coincide. 
\begin{lem}\label{lem1}
For an outer-minimizing surface $\Sigma$, the optimizer $X_{\text{inner}}$ for the Bartnik-Bray inner mass and the HRT surface $X$ for the outer entropy, if they exist, satisfy $A(X_{\text{inner}})\leq A(X).$
\end{lem}
The above lemma follows from the fact that the outer entropy is more restrictive than the inner mass when $\Sigma$ is outer-minimizing. The $X_{\text{inner}}$ on the optimal fill-in $\Omega$ is a locally extremal surface, so it is also a valid HRT candidate on $\Omega\cup_\Sigma\O$ while $X$ might not actually lie on $\Omega$. Since $(\Omega,X_{\text{inner}})$ is a feasible choice for the optimization of the outer entropy, the maximization can only go higher for the outer entropy.

 What about mean-convexity? EW and NR also proved the following lemma~\cite{engelhardt2018decoding,engelhardt2019coarse,nomura2018area}.
\begin{lem}\label{lem2}
For a mean-convex surface $\Sigma$, the HRT surface $X$  for the outer entropy, if it exists,  has area $A(X)\leq A(\Sigma)$.
\end{lem}
 The above two lemmas \ref{lem1},\ref{lem2} imply $A(X_{\text{inner}})\leq A(X)\leq A(\Sigma)$. Note that the optimizer of the inner mass has area less than $A(\Sigma)$ for any surface $\Sigma$ by definition. Had one only required outer-minimizing but not mean-convexity, then one could construct a situation where $\Sigma$ separates the two optimizers in terms of the area, $A(X_{\text{inner}})\leq A(\Sigma) < A(X)$. This can be done, for example, via making $\Sigma$ ``zigzag'' in the null direction and thus the area $A(\Sigma)$ arbitrarily small. Then there is no way that the two optimizers agree. 

From the above lemmas, we see that both conditions are indeed relevant. Nevertheless, it could be that weaker conditions are sufficient for the equivalence. We can state our main result. 
\begin{thm}\label{mainresult}
For an outer-minimizing mean-convex surface, the supremum area of the outer entropy equals to the supremum area of the Bartnik-Bray inner mass.
\end{thm}

\begin{figure}[t] 
  \centering
\includegraphics[width=0.63\linewidth]{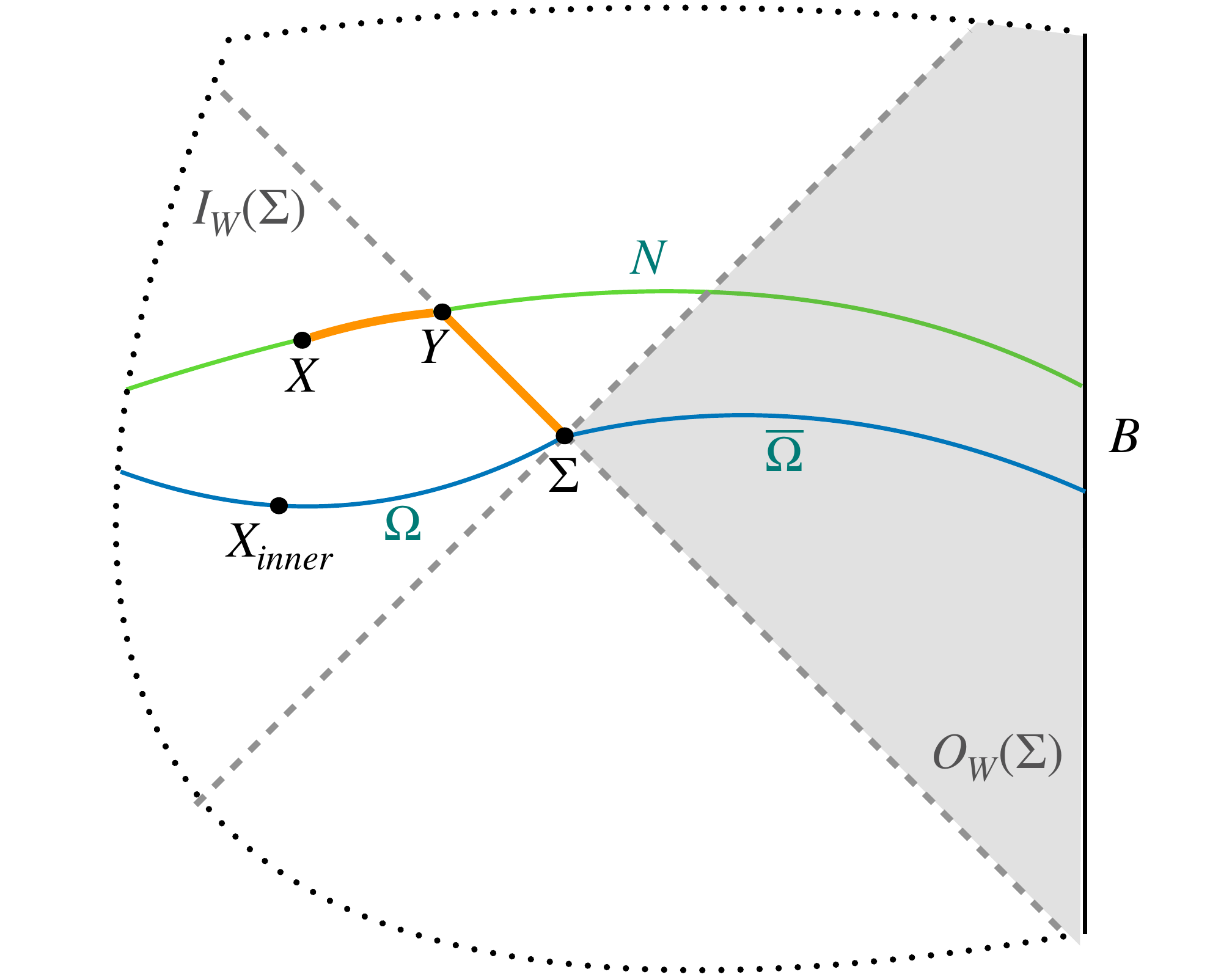}
  \caption{\emph{Portion of a Penrose diagram illustrating the proof of Theorem \ref{mainresult}.} The blue slice $\Omega \cup_\Sigma \O$ consists of an exterior slice $\O$, on which $\Sigma$ is outer-minimizing, and an optimal fill-in $\Omega$, on which $X_{\text{inner}}$ is the minimal surface with maximal area. The green slice $N$ is the maximal slice for the HRT surface $X$ homologous to $\B$. It crosses the future-directed ingoing null congruence from $\Sigma$ at $Y.$ The orange slice $X-Y-\Sigma$ with the data evolved from $\Omega$ also gives a legit fill-in.} 
  \label{proof}
\end{figure}

\begin{proof}

We start by considering the case when the optimizers for both problems exist. Suppose $\Sigma$ is outer-minimizing on $\O$ which is anchored at $\B$, and let the optimal fill-in for the inner mass be $(\Omega,h,K)$ and $X_{\text{inner}}$ is the minimal surface on $\Omega$ with area $A(X_{\text{inner}}).$ Lemma \ref{lem3} implies we only need to consider the HRT surface $X$ inside the inner wedge $D(\Omega)$. If $X$ lies on $\Omega$ then $X,X_{\text{inner}}$ can be identified.  Suppose the HRT surface is realized at some other surface not on $\Omega$. $X$ must have larger area $A(X)\geq A(X_{\text{inner}})$ due to Lemma \ref{lem1}.  $X$ is thus a minimal surface on some Cauchy slice $N$ that does not pass through $\Sigma,$ and we denote the intersection surface between $N$ and the future-directed ingoing null congruence from $\Sigma$ as $Y,$ which is called a representative of $\Sigma$ by EW in~\cite{engelhardt2019coarse}. Since $\Sigma$ is a mean-convex surface $\theta^-(\Sigma)\leq 0$, assuming the NEC, the Raychaudhuri equation implies that $\theta^-\leq 0$ on the whole congruence connecting $\Sigma$ and $Y.$ This gives $A(Y)\leq A(\Sigma)$ and we also have $A(X)\leq A(Y)$ as $X$ is a HRT surface sitting on $N$. Therefore, $X$ is also a minimal surface on the slice $X-Y-\Sigma$~\footnote{Note that we have used a spacelike-characteristic initial data and assumed enough regularity of it for it to be a legitimate fill-in. See more about spacelike-characteristic Cauchy problem with low regularity in\cite{czimek2019spacelike}.}, consistent with Lemma \ref{lem2}, so that $A(X)\leq A(X_{\text{inner}})$, because $X_{\text{inner}}$ is the optimizer for the inner mass. Hence, $A(X_{\text{inner}})=A(X),$ and both $X,X_{\text{inner}}$ can be the optimizer of the outer entropy and the inner mass. This argument\footnote{A similar construction is used in proving the maximin surface~\cite{wall2014maximin} we have been using is the same as the original HRT proposal~\cite{hubeny2007covariant}.} is illustrated in Figure \ref{proof}.

Therefore we can conclude that for an outer-minimizing mean-convex surface, both optimizers of the outer entropy and the Bartnik-Bray inner mass exist or neither exists. Now consider the case when neither optimizer exists. Suppose that the supremum area of $\S$, $A_1$, is strictly larger than $M_{\text{inner}},$ $A_2,\, A_1>A_2.$ Consider now a sufficiently small $\epsilon>0$ such that $A_1-\epsilon>A_2.$ Then there exists an inner wedge configuration for $\S,$ which has the extremal surface area equal to $A_1-\epsilon$ (otherwise, $A_1-\epsilon$ would be the supremum). Therefore, according to above arguments, the supremum area of $M_{\text{inner}}$ should be at least $A_1-\epsilon,$ which gives the contradiction. The other direction of starting with $A_1<A_2$ follows from the same arguments.
\end{proof}

The outer-minimizing condition shows up in different contexts. On the quasilocal mass side, the outer-minimizing condition is not necessary for the inner mass to be well-defined, unlike the outer mass. However, as pointed out by Bray~\cite{bray2004penrose}, it is useful to impose such a condition for the inner mass as well. For example, we can upper-bound the inner mass with the outer mass, assuming the Penrose inequality holds (see the next section). We see an interesting parallel here: \emph{the outer-minimizing condition serves as a necessary condition to guarantee the outer mass being non-trivial, whereas it is meant to make sure the HRT surface exists for the outer entropy via the construction using BNR}.

We finish this section by a side remark about mean-convex surfaces. One might find the mean-convex surface condition ($\mathbf{H}$ being inward spacelike) unnatural and unnecessary for the inner mass and the outer entropy. Although it is not needed for the definition, it serves to establish the equivalence and puts them in the same context as other closely-related quantities that do need mean-convexity.  For example, in the Riemannian setting, mean-convexity means $H\geq 0.$ It is needed for the Bartnik mass $M_B$ to be non-degenerate, because for Bartnik data with negative mean curvature then any extension $(N,h)$ has a horizon~\cite{anderson2019recent}.  There is actually another quasilocal mass proposal, called the Liu-Yau mass, that specifically requires the surface to be mean-convex~\cite{liu2003positivity,miao2015quasi,szabados2009quasi}. It can be considered as a refinement of the Brown-York quasilocal mass~\cite{brown1993quasilocal} in that a positive mass theorem can be proved for the Liu-Yau mass~\cite{liu2003positivity}.

We have proved that the outer entropy is equivalent to the inner mass:
\beq\label{eqn:equalityaf}
M_\mathrm{inner}=\frac12\left(\frac{4\hbar G_N\S(\Sigma)}{\Omega_{n-2}}\right)^\frac{n-3}{n-2}
\eeq
for AF data and
\beq\label{eqn:equalityah}
M_\mathrm{inner}=\frac12\left(\frac{4\hbar G_N\S(\Sigma)}{\Omega_{n-2}}\right)^\frac{n-3}{n-2}+\frac12\left(\frac{4\hbar G_N\S(\Sigma)}{\Omega_{n-2}}\right)^\frac{n-1}{n-2}
\eeq
for AH data.

It's worth noting that BNR already realized that the RHS of (\ref{eqn:equalityaf}) can be used as a quasilocal mass~\cite{bousso2019outer} and (\ref{eqn:equalityah}) was proposed in~\cite{engelhardt2019holographic}. However, like we mentioned in introduction, the RHS is generally not a quasilocal quantity. It becomes quasilocal when one imposes the outer-minimization condition, but then one can use a more straightforward proposal due to Bray and Bartnik.

\section{Implications}\label{sec:implication}

\subsection{Area laws and the monotonicity of quasilocal mass}\label{subsec1}
One important result of EW is that the area laws follow immediately from formulating the black hole entropy in a variational way. More precisely, EW~\cite{engelhardt2018decoding} shows that the area laws associated with the holographic screens\footnote{This is also known as the future trapping horizon. Here actually we need to restrict the holographic screens to be spacelike or null in order to satisfy the minimar condition. This restricted class is also known as the dynamical horizon~\cite{ashtekar2002dynamical,hayward1994general}.} foliated by apparent horizons $\Sigma_0$ follow from the fact that $A(\Sigma_0)=4G\hbar\S(\Sigma_0)$ and the definition of outer entropy \ref{outerentropy}. NR generalizes the area laws to the holographic screens formed by a class of surfaces that are not marginally (anti-)trapped, which includes the case of the event horizon, and also the related second law of the outer entropy~\cite{nomura2018area}. The family of area laws restricted to the spacelike or null part of the holographic screen can be summarized as $A(\Sigma_1)\leq A(\Sigma_2)$, if the outer wedge of $\Sigma_2$ is contained inside the outer wedge of $\Sigma_1$. This is because the constraint space reduces from $\Sigma_1$ to $\Sigma_2$. They are the direct consequences of the monotonicity property built into the definition of the outer entropy \ref{outerentropy}. 

Similarly the Bartnik-Bray quasilocal masses are also defined with such variational formula. The area laws above translate to the monotonicity of the quasilocal mass: given $\Sigma_2$ and $\Sigma_1\subset I_W(\Sigma_2)$ we have~\cite{bartnik1989new,bartnik1997energy,bray2004penrose},
\beq
M_{\text{inner}}(\Sigma_1) \leq M_{\text{inner}}(\Sigma_2).\label{innerlessouter}
\eeq
It holds because a valid fill-in $\Omega_1$ of $\Sigma_1$ can always be turned into a valid fill-in of $\Sigma_2$ by gluing $\Omega_1$ to  $\Sigma_2$ through some initial data set connecting $\Sigma_1$ and $\Sigma_2.$ Similar arguments also apply to the outer mass.

Monotonicity is actually one of the desirable features that a good quasilocal mass proposal should have~\cite{ChristodoulouYau,szabados2009quasi}. It is physically important as it demonstrates the positive mass contribution in a quasilocal way. Also, monotonicity associated with a quasilocal mass is often technically useful. One example is the proof of the RPI by Huisken and Ilmanen~\cite{huisken2001inverse} which uses the monotonicity of the Hawking mass under the inverse mean curvature flow. In short, we see that the holographic screen area laws are mathematically equivalent as the monotonicity of the Bartnik-Bray quasilocal masses. One can view it as a monotonicity associated with processing the geometric data, analogous to the \emph{data processing inequality} in quantum information theory that concerns entropic functions on the data of quantum states\footnote{Data processing inequalities are usually proved using some concavity properties of the entropic functions that are difficult to show. However, for quantities associated with operational meaning, one can always find a variational formulation in terms of some optimization problem, from which the monotonicity is straightforward.}. A canonical example is the monotonicity of relative entropy~\cite{araki1976relative,araki1977relative}, and we shall come back to this in the next subsection.

\subsection{Entropy bounds for the gravitational energy}\label{sec:entropicbounds}

Ever since Bekenstein conjectured that the entropy contained in a finite region is universally upper-bounded by the energy within~\cite{bekenstein1972black,bekenstein1973black,bekenstein1974generalized}, there have been developments of various entropy bounds that relate energy and entropy. One important achievement of such interplay between high energy physics and quantum information is the Quantum Null Energy Condition (QNEC), which is first derived from the Quantum Focusing Conjecture~\cite{bousso2016quantum} and then rigourously proved in~\cite{bousso2016proof,koeller2016holographic,ceyhan2018recovering,balakrishnan2019general}. Recently, Wall argued a universal lower bound of the energy density in classical and quantum field theories~\cite{wall2017lower}, which directly implies the QNEC as a special case. In 1+1 dimensions, consider an ant marching along the line space coordinated by $x$ and we assume the global energy is lower-bounded. At any point $x_0$, the ant wonders what is the minimal amount of the total energy given what she has observed about the matter field configuration $\rho_{\O}$ from $-\infty$ to $x_0$. Mathematically, this is given by the quantity
\beq
\inf_{\rho:\,\tr_{\O}\rho = \rho_\Omega} \int_{x_0}^\infty \langle T\rangle_{\rho_{\O}}\, \dd x 
\eeq
where $\rho$ denotes the purification of the given $\rho_\Omega$ quantum state of the matter and $\rho_{\O}:=\tr_\Omega\,\rho$ is its quantum marginal state that we optimize over;  $T$ is a schematic notation of stress tensor operator of some component (see \cite{wall2017lower} for details), and $\langle T\rangle:=\tr\,T\rho$.  In words, one tries to minimize the total energy over all purifications of the given interior state $\rho_\Omega$. Wall conjectured that this minimal energy is given by the derivative of the von Neumann entropy at $\Sigma$. When generalized to higher dimensions, Wall's ant conjecture formally reads:
\begin{conj}\label{conj:antwall}
Given some partial Cauchy slice $\Omega$ with boundary $\Sigma$, a quantum state $\rho_\Omega$ on $\Omega$ and some unit vector field $X$ on $\Sigma$, the stress tensor $T$ of any quantum field theory satisfies
\beq\label{eq:antwall}
\inf_{\rho:\,\tr_{\O}\rho = \rho_\Omega}\int_{\O}\langle T\rangle_{\rho_{\O}} \,\dd x = \frac{\hbar}{2\pi}\mathcal{L}_X S_\rho(\Omega)|_\Sigma.
\eeq
\end{conj}
\begin{rem}
One motivation for the ant conjecture is that the monotonicity of the minimal energy under all completely
positive maps matches with the strong subadditivity on the right~\cite{wall2017lower}. If one choses $X$ as variations on a null surface, one can obtain an entropic lower bound on $T(X,X)$ (QNEC) by taking the derivative on both sides and using the built-in monotonicity as in the last subsection (cf.~\cite{wall2017lower} and the Appendix of~\cite{bousso2019ignorance} for more details).  
\end{rem}

This version of the ant conjecture will be useful in section~\ref{sec:quantum}. Let us also discuss a slightly different formulation of the ant conjecture studied in~\cite{ceyhan2018recovering,bousso2019ignorance}, which differs from Wall's original proposal in that it concerns the total matter energy. Consider some cut $\Sigma$ on a Killing horizon. Using results of Unruh effect~\cite{bisognano1975duality,wall2012proof} and the data processing inequality of the quantum relative entropy, one can show that $M(\Sigma)$ is lower bounded by~\cite{bousso2019ignorance,wall2017lower}
\beq\label{antineq}
\inf_{\rho:\,\tr_{\O}\rho = \rho_\Omega} \int_{\Omega\cup\O} \langle T\rangle_{\rho}\, \dd x \geq-\frac{\hbar}{2\pi} \mathcal{L}_X D(\rho_{\O}||\sigma_{\O})|_\Sigma
\eeq
where $D(\rho_{\O}||\sigma_{\O}):=\tr\rho_{\O}\log\rho_{\O}-\tr\rho_{\O}\log\sigma_{\O}$ is the relative entropy between two marginal states $\rho_{\O},\sigma_{\O}$ of the global state $\rho$ and the vacuum state $\sigma$ respectively.
 
In this version, the ant conjecture claims the inequality (\ref{antineq}) is in fact an equality~\cite{ceyhan2018recovering,bousso2019ignorance}: 
\begin{conj}\label{conj:ant}
Given a cut $\Sigma$ on some Killing horizon $\Omega\cup_\Sigma\O$, a quantum state $\rho_\Omega$ on $\Omega$ and some unit null vector field $X$ on $\Sigma$, the stress tensor $T$ of any quantum field theory with the vacuum $\sigma$ satisfies
\beq\label{conjant}
\inf_{\rho:\,\tr_{\O}\rho = \rho_\Omega} \int_{\Omega\cup\O} \langle T\rangle_{\rho}\, \dd x  = -\frac{\hbar}{2\pi}\mathcal{L}_X D(\rho_{\O}||\sigma_{\O})|_\Sigma.
\eeq
\end{conj}
\begin{rem}
This second version can be seen as a special case of the original ant conjecture \ref{conj:antwall}. This is because on a Killing horizon, we can represent the energy of the interior state via the modular hamiltonian of the vaccum state (cf. \cite{bousso2019ignorance}). Its general vality in any background is questionable: the data processing inequality of the relative entropy~\cite{araki1976relative,araki1977relative} implies the LHS is positive, which imposes some average positive energy condition on the QFT. Also, the monotonicity of the RHS may not generally hold as opposed to the RHS in (\ref{eq:antwall}).
\end{rem}
\begin{rem}\label{rem:cocycle}
Conjecture \ref{conj:ant}(and \ref{conj:antwall}) has been proven by Ceyhan and Faulkner~\cite{ceyhan2018recovering} when $X$ is restricted to variations on a Rindler horizon in Minkowski spacetime. They use a particular family of purifications of the given marginal state. This is a one parameter family of unitaries acting on $\rho_{\O}$ known as the Connes cocycle flow. In the limit of the parameter approaching the infinity, one achieves equality in (\ref{conjant}). 
\end{rem}

The above mentioned results concern the von Neumann entropy of the matter fields and its stress tensor. It is well known that, in the absence of matter fields, a covariant characterization of the local gravitational energy is forbidden in general relativity~\cite{szabados2009quasi,misner2017gravitation} and the quasilocal mass is our best alternative. As a quasilocal mass, the inner mass is actually not as physically motivated as the outer mass, but we could use the equivalence we just established to look for relations between the outer mass and the outer entropy. It is obvious that Penrose inequality plays an important role in this.

Assuming the Penrose inequality, one can show that for outer-minimizng $\Sigma$~\cite{bray2004penrose}\footnote{The inequality is reminiscent of the weak duality property between the primal and the dual convex optimization programs, which again are often used in quantum information theory. }
\beq
M_{\text{outer}}(\Sigma)\geq M_{\text{inner}}(\Sigma) .\label{weakduality}
\eeq
It simply follows from the Penrose inequality (\ref{PI}), which requires that, even in the worst case,  the maximal irreducible mass filled into $I_W(\Sigma)$ is always upper-bounded by the minimal ADM mass associated with the extensions in $O_W(\Sigma).$  In turn, one can then bound the outer mass with the outer entropy $\S$ for AF data:
\beq
\inf_{(\O,h,K)\in\PMS_\Sigma} M(\O,h,K) \geq M_\mathrm{irr}(4\hbar G_N\S(\Sigma))\label{graviant}
\eeq
where the LHS is the infimum of the total energy while holding a portion of spacetime fixed, and the RHS is an entropy term which summarizes (\ref{eqn:equalityaf},\ref{eqn:equalityah}).

One could think of (\ref{graviant}) as a gravitational version of (\ref{antineq}), albeit the entropy bound looks somewhat different, such as that no derivative is taken on the outer entropy. As opposed to the matter case, we do not yet fully understand\footnote{In the perturbation theory regime, however, EW gives a proposal of \emph{simple entropy} as the boundary dual of the outer entropy~\cite{engelhardt2019coarse}.} the quantum definition the outer entropy, so the definition in (\ref{formalouter}) cannot work independently without the bulk. Also, the quasilocal mass has never been studied beyond the framework of general relativity. It would be insightful to interpret (\ref{graviant}) from the boundary field theory perspective in the holography context.

There are also various localized Penrose inequalities that lower bound other quasilocal masses with the irreducible mass~\cite{miao2009localized,ho2013localized,lu2017minimal,chen2018quasi,wang2019localized,alaee2019geometric}, and all of them can be potentially turned into such entropy bounds. However, their physical meanings are  more obscure, so we do not consider them here.

\subsection{The gravitational ant conjecture}

 In case of (\ref{weakduality}, \ref{graviant}) being saturated, it resembles Wall's ant conjecture.  Here we propose a \emph{gravitational ant conjecture}: given a cut $\Sigma$ on a hypersurface and the induced Bartnik data, the infimum  of the total mass over the extensions is given by the irreducible mass of the outer entropy.
\begin{conj}\label{conj:classical}
Given a Bartnik darta set $(\Sigma,\gamma,H,k,\omega^\perp)$ associated with a codimension-two surface $\Sigma$, we have
\beq
\inf_{(\O,h,K)\in\PMS_\Sigma} M(\O,h,K) = M_\mathrm{irr}(4\hbar G_N\S(\Sigma)).\label{conjgraviant}
\eeq
\end{conj}

\begin{rem}
The gravitational ant conjecture itself is a purely geometric statement. It resembles Conjecture \ref{conj:ant} and also Conjecture \ref{conj:antwall} (see more discussion in section~\ref{sec:quantum}). 
\end{rem}
\begin{rem}\label{rem:compute}
Although the outer entropy is computed with replacing the original geometry by some optimal fill-in, we only vary over the extensions while fixing the interior in (\ref{conjgraviant}).
\end{rem}
Although the entropy bounds in (\ref{conjgraviant}) and (\ref{eq:antwall}) look apparently different, one important indication of this conjecture is that both sides enjoy the monotonicity as we discussed in section \ref{subsec1}. It is not yet known what are the general conditions on the Bartnik data for the above equalities to hold, and it's plausible that one needs to further constrain the data $(\Sigma,\gamma,H,k,\omega^\perp)$ to prove it. Nevertheless, we do know the conjecture can be realized by the (AdS-)Schwarzschild spacetime due to the rigidity part of the Penrose inequality conjecture. For instance, if $\Sigma$ encloses the horizon in the Schwarzschild data, then the inner mass matches with the outer mass.  In the Riemannian AF setting, it turns out for the Bartnik data associated with an horizon, one can construct a one-parameter family of spacetime metrics such that the equality in (\ref{conjgraviant}) is achieved in the limit of the parameter going to the infinity.

\begin{figure}
  \centering
\includegraphics[width=0.85\linewidth]{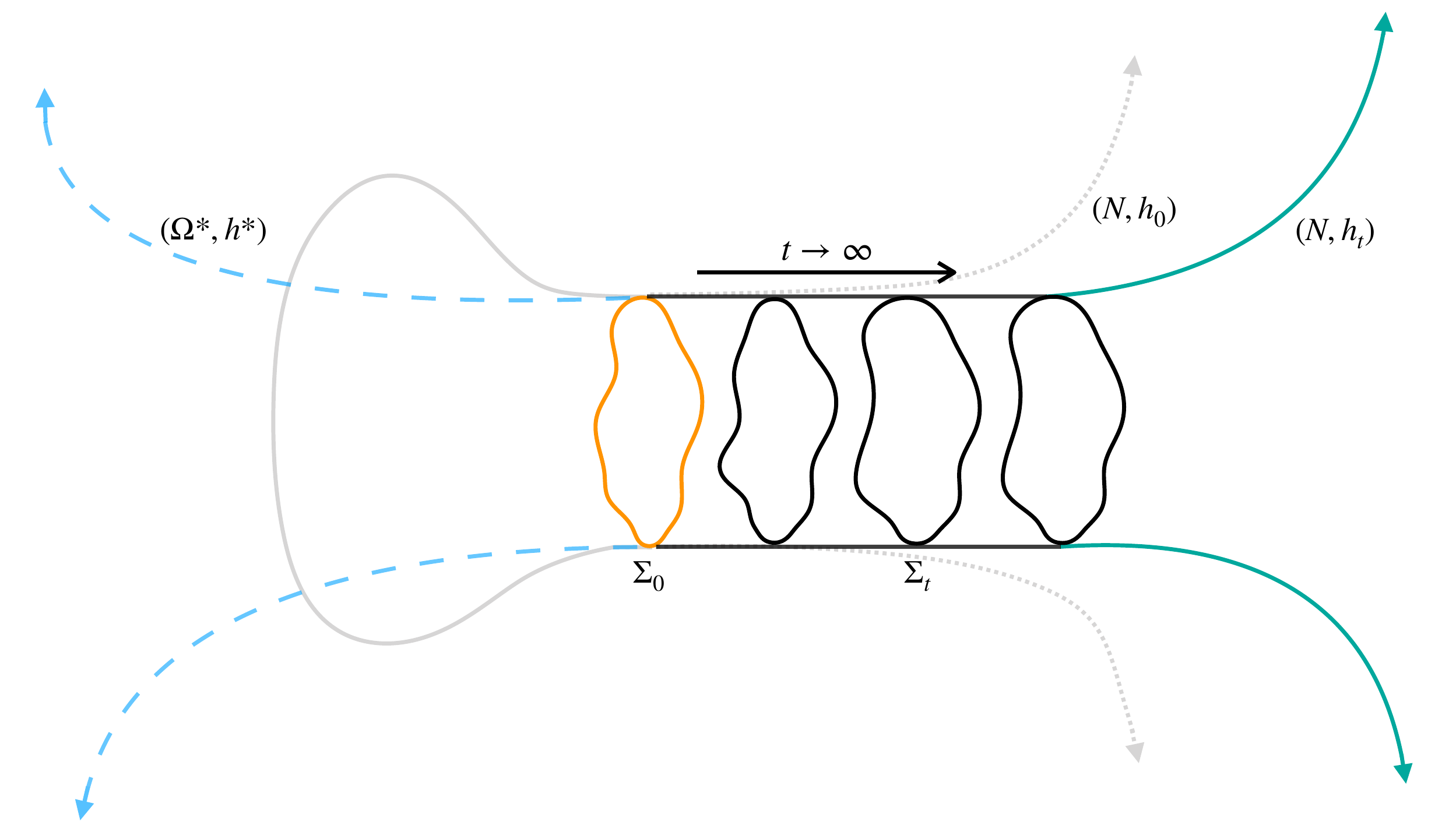}
  \caption{\emph{Bray's conformal flow of metrics.} The grey line represents the original data set $(N,h_0)$ which contains a horizon $\Sigma_0$ in orange. The solid line depicts the new initial data. The interior region is kept unchanged. The flow $h_t$ conformally transform the metric outside the horizon, resulting in an outward flow of the horizon $\Sigma_t$ with the area unchanged, thus depicted as the black cylinder region. The exterior region outside the horizon $(N,h_t)$ in green becomes arbitrarily close to the Schwarzschild metric for large enough $t$. $\Sigma_0$ remains outer-minimizing in $(N,h_t)$. The dashed blue region is a fill-in $(\Omega^*,h^*)$ that connects to another AF end with $\Sigma_0$ being the wormhole neck, whose area gives the outer entropy. The resulting spacetime is \emph{not} the time-symmetric initial data for the Schwarzschild spacetime. } 
  \label{conformal}
\end{figure}

This construction is due to Bray in his proof of the RPI~\cite{bray2001proof} (cf. Figure \ref{conformal} for an illustration). Bray constructs a conformal flow of metrics $h_t:= u_t^n h_0$, where the conformal factor $u_t^n$ fixes the interior part of the initial but stretches the exterior metric. This flow satisfies the property that the area of the horizon $\Sigma_t$, defined as the outermost minimal surface, is invariant. Most importantly, the ADM mass monotonically decreases under the flow and the exterior data tends to the spatial Schwarzschild metric. Hence, the Penrose inequality is established. Note that the original horizon $\Sigma_0$, despite no longer being the outermost at $t>0$, remains outer-minimizing in $(N,h_t)$, and the Bartnik data is in fact matched at $\Sigma_0$. Therefore, a by-product of this conformal flow method is that the Bartnik-Bray outer mass of any Riemannian Bartnik data $(\Sigma,\gamma,H)$ is given by the irreducible mass (\ref{irr})\footnote{This argument has the advantage that it works for the spacetime with multiple black holes~\cite{privatebray}. However, this construction does not give any results on the original Bartnik mass $M_B$. The fact that the Bartnik mass of a horizon is the irreducible mass has been proved by Mantoulidis and Scheon~\cite{mantoulidis2015bartnik}, who use a ``collar extension'' construction that works for both the Bartnik mass and the Bartnik-Bray outer mass.}. The RHS of (\ref{conjgraviant}) is trivially given by the irreducible mass of the horizon, which one can also directly infer from the result of EW~\cite{engelhardt2018decoding} (cf. section \ref{sec:discuss}). Therefore the Penrose inequality and (\ref{weakduality}, \ref{graviant}) are saturated in the limit of $t\rightarrow \infty.$ Furthermore, the data in the limit $(N,h_\infty)$ satisfies that the outer mass of any surface enclosing $\Sigma_0$ is the same as $M_{\text{outer}}(\Sigma_0)$. This implies that the exterior region outside $\Sigma_0$ contributes nothing to the ADM mass of $(N,h_t).$

It is worth noting the structural similarity between the above conformal flow of metrics concerning the gravitational energy, and the Connes cocycle flow used by Ceyhan and Faulkner in proving Wall's ant conjecture concerning the matter energy (cf. \ref{rem:cocycle}): generally there are no minimizers of the LHS in both problems; both use a one-parameter family of flow that acts on a part of the field configurations, and the flow yields the equality condition as the parameter goes to the infinity. Also, when we focus on the far region in the conformal flow of metrics, the limit is approached exponentially in the parameter~\cite{bray2001proof}.
\beq\label{eq:expoentialflow}
h_t := u_t(x)^n h_0, \quad\quad \lim_{x\rightarrow\infty}u_t(x)=e^{-t}.
\eeq
Similarly, the limiting state that yields the ant conjecture is also approached exponentially in the cocycle flow. Although the flows are completely different objects in the two contexts, it is tempting to conjecture a concrete connection between them. One could hope to establish a duality between them in the framework of AdS/CFT correspondence. However, the main obstacle is that it is not yet known if the conformal method can work in the hyperbolic setting, so the hyperbolic Penrose inequality is still an open problem in mathematical relativity~\cite{mars2009present}. 

Bousso et al~\cite{bousso2019ignorance} also proposed a classical bulk dual to the Connes cocycle flow on the boundary quantnum field theory. They consider a null hypersurface with a codimension-one cut. Their proposal, called the \emph{left stretch}, is to rescale the affine parameter on a null hypersurface on the left of the cut while keeping the right intact, and then glue them back and treat the new parameter as affine. We would like to point out that this is equivalent to a conformal metric flow on the left region while keeping the right region intact. The above conformal flow equation (\ref{eq:expoentialflow}) implies that in the near-boundary region, rescaling the affine parameter by $e^{t}$ achieves the same effect as Bray's flow. In holography, such exponential behaviour of the bulk metric flow in the near-boundary region can perhaps be universally identified as the cocycle flow in the boundary quantum state. The resemblance could indeed be more than a coincidence.

\section{Application: the small sphere limit}\label{sec:ssl}
Due to the equivalence established in Theorem \ref{mainresult}, the BNR algorithm computes both the outer entropy and the inner mass. 
We would like to apply the algorithm to calculate their small sphere limits. The small sphere limit serves as an important sanity check for a valid quasilocal mass proposal~\cite{szabados2009quasi}. From physical arguments and evidences of other quasilocal mass proporsals~\cite{horowitz1982note,bergqvist1994energy,brown1999canonical,szabados2009quasi,yu2007limiting,chen2018evaluating}, we learn that the small sphere limit, evaluated along lightcone cuts shrinking towards a point $p$ along direction $e_0$, is given by the stress tensor $T(e_0,e_0)|_p$ in the leading order\footnote{For vacuum spacetime where the stress tensor vanishes, we expect the leading order to be given by the Bel-Robinson tensor. We leave the vauum case calculations to future works.}.  The lightcone cut construction, parameterized by the lightcone vertex $p$ and any future timelike unit vector $e_0,$ is a standard way to evaluate the small sphere limits introduced by Horowitz and Schmidt in studying the Hawking mass~\cite{horowitz1982note}. Let's denote the lightcone cuts as $\Sigma_l$ and the small sphere limit can be extracted by computing
\beq
\lim_{l\rightarrow 0}l^{-(n-1)}M_{\text{inner}}(\Sigma_l).
\eeq
 Sufficiently small lightcone cuts are guaranteed to be mean-convex and outer-minimizing so we can apply the algorithm. The algorithm entails solving a polynomial constraint equation parameterized by the Bartnik data on $\Sigma.$ Since it is only shown to be optimal for spherical untrapped surfaces~\cite{nomura2018area,bousso2019outer}, we assume that the lightcone cut $(p,e_0)$ is approximately spherically symmetric including the leading perturbation order due to curvature\footnote{\label{fn:ssl}Note that we did not assume the spacetime is spherically symmetric as in~\cite{nomura2018area}, so we need to start with the full BNR algorithm provided in BNR~\cite{bousso2019outer}. Nevertheless, we find that the constraint equations reduce to the one in~\cite{nomura2018area}. }, which is the leading order we care about in the small sphere limit. 

We perturbively expand the null frame variables on the lightcone cuts in Riemann Normal Coordinates and input them to the BNR algorithm, while making sure all the prerequisite conditions are satisfied. The small sphere limit of the outer entropy at the leading order is given by 
\beq
\S(\Sigma_l)=\frac{\Omega_{n-2}l^{n-2}}{4G_N\hbar}\left(\frac{2l^2\Omega_{n-2}G_N T(e_0,e_0)|_p}{n-1}\right)^{\frac{n-2}{n-3}},
\eeq
and $4G_N\hbar\S(\Sigma_l)$ gives the supremum area of the HRT surface.

We see that the outer entropy at the small sphere limit is directly characterized by the stress tensor. In turn, we also obtain the limit of the inner mass
\beq
\lim_{l\rightarrow 0}l^{-(n-1)}M_{\text{inner}}(\Sigma_l) =\frac{1}{2l^{n-1}}\left[l^{n-2}\left(\frac{2l^2\Omega_{n-2}T(e_0,e_0)|_p}{n-1}\right)^{\frac{n-2}{n-3}}\right]^\frac{n-3}{n-2} =\frac{\Omega_{n-2} T(e_0,e_0)|_p}{n-1},
\eeq
which is exactly the result we expect from volume $\frac{\Omega_{n-2}}{n-1}$ times the matter energy density $T(e_0,e_0)|_p$. It matches with the non-vacuum small sphere limits of other quasilocal masses in four dimensions~\cite{szabados2009quasi} and higher dimensions~\cite{wang2020small}. Note that this result holds for both the AF case (\ref{eqn:equalityaf}) and the AH case (\ref{eqn:equalityah}), because the second term with higher power in (\ref{eqn:equalityah}) is irrelevant at the leading order. 

The review of the BNR algorithm and the detailed calculations are provided in the Appendix \ref{appA} and \ref{appB} respectively. To our knowledge, this result is new. The small sphere limit of the Bartnik mass is not known before except in the Riemannian case~\cite{wiygul2018bartnik}. We believe that the spherical symmetry for the lightcone cut is not needed for this result to hold, as in the case for other quasilocal mass proposals.  It is worth looking into how the technical difficulties can be overcome to prove the optimality of the BNR algorithm or otherwise refine it.

\section{The semiclassical case}\label{sec:quantum}
So far all we have discussed concern the classical gravity. Now we briefly sketch the some possibilities to generalize the gravitational ant conjecture to a quantum statement in the semiclassical setting. It's a fruitful approach to understand semiclassical quantum gravity via replacing the area by the generalized entropy in a classical geometric statement and extracting its implications. In particular, we add to the area, which represents the gravitational coarse-grained entropy, the contribution of the fine-grained von Neumann entropy of the matter on the exterior region $\O$. The \emph{generalized entropy} is defined as
\beq\label{boussotrick}
S_\mathrm{gen}=\frac{A_\mathrm{gen}(\Sigma)}{4G\hbar} =  \frac{A(\Sigma)}{4G\hbar}+S_\rho(\O).
\eeq
The discovery of the QNEC marks the culmination of this line of thought~\cite{bousso2016quantum}. The same idea can be applied to our gravitatonal ant conjecture. Several related ideas have been proposed, such as the quantum EW coarse-graining~\cite{bousso2019ignorance} and the quantum Penrose inequality~\cite{bousso2019quantumprl,bousso2019quantumprd}. We can gain some insights from these works. Note that the original ant conjecture conerning the matter sector is a quantum statement, so a plausible quantum version of our conjecture can be obtained by combining the gravitational (Conjecture \ref{conj:classical}) and matter (Conjecture \ref{conj:antwall}) sectors, in the spirit of~(\ref{boussotrick}).
Following~\cite{bousso2019ignorance}, one can start by defining the generalized outer entropy as maximizing the generalized entropy over the interior data while holding the exterior fixed:
\beq
\S_\mathrm{gen}(\Sigma):=\sup_{(\Omega,h,K,\rho_\Omega)}S_\mathrm{gen}(\Sigma).
\eeq

Since the conjecture~\ref{conj:classical} essentially follows from the Penrose inequality, one might firstly attempt to upgrade its RHS by the quantum Penrose iequality (QPI)~\cite{bousso2019quantumprl}, where one substitutes $\S$ with $\S_\mathrm{gen}$. However, the proposed QPI only works for lightsheets attached to the  apparent horizon (quantum MOTS), which extends to the singularity. It makes their QPI inapplicable to our case as we need a cauchy slice that extends to infinity to define the total mass. 

Alternatively, if we only consider $\Sigma$ being a quantum MOTS, the same idea as in~\cite{bousso2019ignorance} can be used to construct the minimal energy extension. In~\cite{bousso2019ignorance}, it is argued that the optimal state that achieves Wall's ant conjecture (\ref{conjant}) has all the energy concentrated in a shock at $\Sigma$. The same should apply to the mass minimization over the exterior in \ref{conj:classical} when we include the state of mass field in the data. For a quantum marginally trapped surface, the strength of the energy shock has the exactly right magnitude to shift the classical expansion on $\Sigma$ to zero. The proofs due to Bray~\cite{bray2001proof} and Mantoulidis-Schoen~\cite{mantoulidis2015bartnik} in the Riemannian case suggest that the Bartnik mass of an apparent horizon should also be equal to its irreducible mass in the general case, that is $A(\Sigma) = 4G\hbar\S(\Sigma)$. Therefore, we have a semiclassical conjecture in flat spacetime (and we do not repeat the AdS version here):
\beq\label{eq:quantumant}
\inf_{(\O,h,K,\rho_{\O})} M(\O,h,K,\rho_{\O}) = \frac12\left(\frac{A(\Sigma)}{\Omega_{n-2}}\right)^\frac{n-3}{n-2}+\frac{\hbar}{2\pi}\mathcal{L}_X S_\rho(\Omega)|_\Sigma
\eeq
where in the LHS we add in the data the quantum state of the matter field, which is consistent with the rest geometric data in terms of the semiclassical Einstein equation\footnote{In the semiclassical regime, the dominant energy contribution comes from the ADM mass, but one should also include perturbation due to quantum state that doesn't backreact, $\delta M = \int_{\Omega\cup\O} T \,\dd x^{n-1}$.}, and the von Neumann entropy term on the RHS is the matter energy contribution due to the shock. Therefore, we can think of (\ref{eq:quantumant}) as adding up two ant conjectures (\ref{eq:antwall},\,\ref{conjgraviant}) together. We would like to stress again that one does not vary the interior geometry in (\ref{eq:quantumant}) (cf. Remark \ref{rem:compute}). One performs the interior maximation just to compute $\S$ but we keep the original interior data and state $\rho_\Omega$ when considering the minimization of mass over the exterior data. 

The above conjecture hints at an alternative proposal for the QPI:
\beq\label{eq:qpi}
M(\O,h,K,\rho_{\O}) \geq \frac12\left(\frac{A(\Sigma)}{\Omega_{n-2}}\right)^\frac{n-3}{n-2}+\frac{\hbar}{2\pi}\mathcal{L}_X S_\rho(\Omega)|_\Sigma.
\eeq

As opposed to the QPI proposed in~\cite{bousso2019quantumprl}, we still use the original area but we add a reminder term on the RHS. In~\cite{bousso2019quantumprd}, the case of using the bare area of the quantum extremal surface is also considered, and authors argue that it is not enough to compenstate for the negative energy in the Boulware-like state. In our case, note that this entropy derivative term doesn't have a definite sign as opposed to the relative entropy, therefore it would be interesting to check if this remainder term can help compensate for the negative energy\footnote{It is plausible that Wall's ant conjecture is incompatible withe the Boulware-like state as the total matter energy is assumed to be positive~\cite{wall2017lower}. }. We leave this to future study.

In short, it is highly suggestive that the ant conjectures, QNEC, the quantum coarse-graining and the quantum Penrose inequality are all interwined notions where one embeds in the limit of another. Indeed some of the relations are discovered in earlier works. It's worth studying if the gravitational ant conjecture fits in as well.

\section{Discussion}\label{sec:discuss}

For an outer-minimizing mean-convex surface, we have shown that its Bartnik-Bray inner mass is equivalent to its outer entropy proposed by Engelhardt and Wall. Though motivated by completely different problems, these authors arrived at the same optimization construction which manifests their monotonicity. In hindsight, their equivalence suggests these ideas could be profoundly related to each other. By leveraging the Penrose inequality, we conjecture that the minimum global energy, while hold some interior region fixed, is given by its outer entropy, parallel to the ant conjecture due to Wall concerning the matter sector. The conjecture itself as a geometric statement is of independet mathematical interest. Proving the conjecture hinges on a good understanding of the Bartnik mass, which is known to be a difficult problem~\cite{anderson2019recent}. We hope progress can be made by restricting to simple Bartnik data sets. In fact, we know the conjecture is true in some simplified situations. There is also qualitative evidence in support of it when considering the proofs that establish both ant conjectures in these special cases. In particular, the Bray's conformal flow of metrics, which proves the gravitational ant conjecture for an apparent horizon in the Riemmannian case, is structurally analogous to the Ceyhan-Faulkner's cocycle flow proof of the any conjecture in the null case. This analogy can perhaps be solidified in the holographic setting when we consider the Ceyhan-Faulkner's cocycle flow being implemented on the boundary QFT. The exponential behaviours of both flows match at the boundary provide strong evidence for the duality. We've also pointed out several possibilities to upgrade the gravitational ant conjecture into a semiclassical quantum statement and proposed a quantum Penrose inequality. We hope new insights can be gained when combining the ant conjectures of both gravitational and matter sectors. We leave these investigations to future works.

It is a subtle issue to characterize energy in general relativity. Globally, we have a good understanding of the total mass of the spacetime, whereas locally there is much more trouble. It is forbidden to finely resolve the energy content locally, and the best one can do is to obtain a coarse quasilocal quantity of a closed domain. This difficulty is essentially due to the equivalence principle~\cite{szabados2009quasi,misner2017gravitation}. Of course, there are other alternatives such as the gravitational pseudo-tensors, which unfortunately break the general covariance. So far, various quasilocal mass proposals represent our best understanding of gravitational energy in general relativity. The gravitational ant conjecture gives us a qualitative interpretation of the quasilocal mass. If we compare the original ant conjecture (\ref{conjant}) to the gravitational one (\ref{conjgraviant}), there are two main differences: 1. the outer entropy is a coarse-grained entropy as opposed to the relative entropy which is fine-grained; 2. the energy in the matter case can be localized whereas it is not possible for the gravitational quasilocal mass. It is plausible that the two points are correlated, it suggests that because of the coarse-grained nature implied from the RHS of (\ref{conjgraviant}) rather than a fine-grained entropy, one cannot write the gravitational energy as an integral over local energy densities like in (\ref{conjant}). To put the intuition on more solid grounds, a thorough understanding of the outer entropy holograhic boundary dual is indispensable.

Classically, we would like to understand how the outer entropy can be computed for an arbitrary Bartnik data, so that we can compute the small sphere limit in vacuum for example. The BNR algorithm is promising but may not be optimal in general. The problem is perhaps easier in the Riemannian setting. So far we've been working in the spacetime setting where the initial data is given by $(N,h,K)$. In the Riemannian setting, we set $K=0$ and it corresponds to a time-symmetric slice embedded in the spacetime. The DEC reduces to a condition on the Ricci scalar of $(N,h): R\geq 0$ for AF data and $R\geq -(n-1)(n-2)$ for AH data.  Various perspectives of the Bartnik mass and fill-in problem are much better understood in the Riemannian setting than in the spacetime setting~\cite{anderson2019recent}. For example, the RPI is proven in dimensions less eight~\cite{huisken2001inverse,bray2001proof,bray2009riemannian} so the upper bound of the outer entropy stated in (\ref{graviant}) holds as a theorem in the Riemannian setting. Therefore, it would be fruitful to consider the static version of the outer entropy that is measured by the area of the Ryu-Takanayagi (RT) surface~\cite{ryu2006holographic,ryu2006aspects}. One interesting problem is whether a similar construction can be found for the outer entropy of a surface $\Sigma$ on a time-symmetric slice. Geometrically, this corresponds to the Riemannian version of the optimal fill-in problem, where the set of permissible fill-in are restricted to initial data with vanishing second fundamental form. A trivial example is $\Sigma$ being a horizon. The maximal area cannot be larger than the horizon area and since the horizon itself is a minimal surface, the inner mass is just given by the irreducible mass. This is consistent with the EW result as the apparent horizon on a time-symmetric slice $(N,h)$ is locally extremal and outer-minimizing in $(N,h)$, so $\Sigma$ is also the HRT surface with respect to the boundary. Another example is a round sphere $S$ in a spherically symmetric spacetime. The outer mass of such a sphere $S$ can be shown to match the standard Misner-Sharp mass for round spheres using the RPI~\cite{szabados2009quasi}. Since the minimal extension is the Schwarzschild metric outside $S$, the inner mass and outer masses coincide, and then the outer entropy can be calculated using the Misner-Sharp mass. Also, we already know that the Bartnik data always admits a fill-in with minimal surfaces when the mean curvature is below some threshold value~\cite{jauregui2013fill}. Therefore, it is an easier task to first understand the Riemannian problem before tackling the general spacetime case.

\acknowledgments

I am grateful to Aron Wall for his detailed comments on the first draft of this work. I also thank Raphael Bousso for discussions on the outer entropy, and Hubert Bray, Alessandro Carlotto, Gary Horowitz and Jeffrey Jauregui on quasilocal mass. This work is supported by the Swiss National Science Foundation via the National Center for Competence in Research ``QSIT", and by the Air Force Office of Scientific Research (AFOSR) via grant FA9550-16-1-0245.

\bigskip

\appendix
\section{The BNR algorithm}\label{appA}
The main idea to construct the optimal spacetime for the interior of $\Sigma$ followed by both EW~\cite{engelhardt2019coarse} and BNR~\cite{nomura2018area,bousso2019outer} is to use the characteristic initial value formalism which guarantees an unique spacetime evolved from the initial data glued to $\Sigma$ in the interior. To achieve the optimality, such data is put in by hand on a null hypersurface $N_{+}$ emanating from $\Sigma$ in the direction of $-\ell^+,$ and they are constrained by the following set of equations
\beq
\begin{aligned}
  \nabla_+ \theta^+ &= -\frac{1}{n-2} \theta^{+2} 
    - \sigma^{+2} - 8 \pi G_N\,R_{++},  &(\text{Raychaudhuri})\\
  q_a^{\;\;b} {\cal L}_k \omega^+_b &= -\theta^+ \omega^+_a 
    + \frac{n-3}{n-2}{\cal D}_a \theta^+ - {\cal D}_b \sigma^{+b}_a 
    + 8 \pi G_N\,T_{a+}, &(\text{Damour-Navier-Stokes}) \\
  \nabla_+ \theta^- &= -\frac{1}{2}{\R} - \theta^+ \theta^-
    + \omega^{+2} + {\cal D}\cdot \omega^+ + 8 \pi G\,T_{+-} + \Lambda. &(\text{Cross-focusing})
\end{aligned}
\label{eq:constraintk}
\eeq
where $\D$ is the covariant derivative on $\Sigma$.

Since the small light cone cuts are mean-convex surfaces, we shall follow the algorithm proposed by BNR \cite{bousso2019outer}. We only sketch their proposal here and one shall refer to~\cite{bousso2019outer} for more details. We set the cosmological constant $\Lambda=0$ in order to be comparable with the small sphere limits of other quasilocal masses in the literature. BNR first use the constraint equations to locate a marginally trapped surface $\mu$, and then the EW arguments can be used to show that the HRT surface, if exists, has area equal to $A(Y)$. By choosing the stress tensor and shear to vanish for the sake of optimality, the above constraint equations in dimensions $n\geq 6$\footnote{ Note that the above equations hold for dimensions $n\geq 6$, and similar equations are stated in \cite{bousso2019outer} for $n=3,4,5$ separately. For simplicity, we only discuss $n\geq 6$ here and results in the other dimensions are basically the same.
} reduce to
\beq
\begin{aligned}
\left( \theta^+\theta^- - \rho - \epsilon_{1} - \epsilon_{2} 
    - \epsilon_{3} - \epsilon_{4} - \epsilon_{5} \right) \xi^{n-1}+\rho\xi^{2} + \epsilon_{1}\xi^{3} + \epsilon_{2}\xi^{4} 
    + \epsilon_{3}\xi^{n} + \epsilon_{4}\xi^{n+1} + \epsilon_{5}\xi^{2n-2} = 0,
\label{eq:polynomial}
\end{aligned}
\eeq
where 
\beq
\begin{aligned}
\rho & =-\frac{1}{2}\frac{n-2}{n-3}\,\R, \\
\epsilon_{1} & =\frac{n-2}{n-4}\left(\Box\log\theta^+ -|{\cal D}\log\theta^+ |^{2}\right), \\
\epsilon_{2} & =2\frac{n-2}{n-5}|{\cal D}\log\theta^+ |^{2},\\
\epsilon_{3} & =-(n-2)(\D\cdot\omega-\Box\log\theta^+ -(n-2)\omega\cdot\D\log\theta^+ +(n-2)| \D\log\theta^+ |^2),\\
\epsilon_{4} & =-\frac{n(n-2)}{2} \left(\omega\cdot\D\log\theta^+ - \Box\log\theta^+\right),\\
\epsilon_{5} & =-\frac{n-2}{n-1}|\omega - \D\log\theta^+ |^2,
\end{aligned}
\label{eq:coeffi}
\eeq
where $\Box={\cal D}\cdot{\cal D}$  and all the data is evaluated on $\Sigma$ so we omit the arguments of the variables. The parameter
\beq
\xi (\nu, x^i)^{-1}:=\frac{\theta^+(\nu)}{\theta^+} = \left(+\frac{\nu(x^i)\theta^+}{n-2}\right)^{-1}
\eeq
 is measuring how the outer expansion changes with respect to value on $\sigma$ along null generators flowing down $N_+$ parameterized by $\nu.$ Note that the constraint eqaution (\ref{eq:polynomial}) is gauge-dependent. If we rescale the null generators while keeping their inner product,
 \beq
 \ell^\pm \rightarrow \exp(\pm\Gamma)\ell^\pm, \,\nu\rightarrow \exp(-\Gamma)\nu,
 \eeq
the following quantities will change accordingly,
\beq
\theta^\pm\rightarrow \exp(\pm\Gamma)\theta^\pm,\,\omega^\pm\rightarrow\omega^\pm\pm\D\Gamma.
\eeq 

We see that $\xi$ is invariant but the zeros $\xi_0$ of (\ref{eq:polynomial}) might change. In general, $\xi$ depends on the transverse directions $x^i$ as well, but we would like to choose a gauge such that it is independent of $x^i,$ 
\beq\label{condition1}
\xi_0 = 1+\frac{\nu\,\theta^+}{n-2} \;\;\;\;\;\;  \;\;\;  \;\;\;  \;\;\;  \;\;\;  \;\;\;  \;\;\;    (\text{Condition 1})
\eeq
for some $\nu$ independent of $x^i$. Condition 1 guarantees that we indeed obtains a marginally trapped surface $\Sigma_0$ at $\nu(\xi_0).$ If in addition, we have 
\beq
\partial_-\theta^+ (Y_0) < 0\;\;\;\;\;\;\;\;\;\;\;\;\;\;\;\;\;\;\;\;\;\;\;\;\;\;\;\;  (\text{Condition 2})
\eeq\label{condition2}
then we know $\Sigma_0$ is a minimar surface, so we can follow EW to construct the HRT surface. We do not need to explicitly construct the HRT surface, as all we want is the outer entropy measured by the HRT surface:
\beq\label{hrtarea}
\S = \frac{A(X)}{4G_N\hbar}=\frac{A(\Sigma_{0})}{4G_N\hbar}=\frac1{4G_N\hbar}\int_\sigma \xi_0^{n-2}\,\dd \sigma
\eeq
where the second equality is due to the fact that $X$ is obtained from $\Sigma_0$ through a flow on a stationary null hypersurface as constructed by EW. Since $\xi_0>1$, we see that 
\beq
\S(\sigma)<\frac{1}{4G_N\hbar}\int_\sigma \dd\sigma=\frac{A(\sigma)}{4G_N\hbar}
\eeq
 as claimed. Practically, to execute the algorithm, one can start by choosing some appropriate gauge for the null variables and compute $\xi_0$. If condition 1 is not satisfied, one needs to tune the gauge accordingly such that condition 1 can be satisfied. Then one also needs to check condition 2 so that we know a HRT surface exists following EW. Otherwise, the BNR algorithm does not apply to the surface $\sigma$ chosen.

Note that if we assume the validity of the HRRT prescription, we no longer need the holographic duality in order to define and evalute the outer entropy, and we can work in a spacetime that is aymptotically flat. It would be interesting to ask what the outer entropy tends to as $\sigma$ apporaches the spatial infinity in an aymptotically flat manifold and if it has any relation with the ADM mass.

\section{The outer entropy and the bulk stress tensor}\label{appB}
The lightcone cuts $\{S_l\}_l$ parameterized by $(p,e_0)$ are defined as the following~\cite{horowitz1982note}. Let $L_p$ denote the future-directed lightcone generated by null generators $\ell^+$ parameterized by affine parameter $l$. We pick a future-directed timelike unit vector $e_0$ and normalized $\ell^+$ at $p$ by 
\beq
\langle e_0,\ell^+ \rangle =-1.
\eeq
The lightcone cut is the family of codimension-two surfaces $S_l$ define as the level sets of $l$ on $L_p$. The ingoing null generators on $L_p$ are denoted as $\ell^-$ and they are normalized by
\beq
\langle \ell^-,\ell^+ \rangle =-1.
\eeq
The small sphere limit along lightcone cuts are given taking $l$ to zero. This is a canonical way to evaluate the small sphere limits of quasilocal mass. 

We shall evaluate some intrinsic and extrinsic geometric quantities that are needed as the input to the BNR algorithm. We do this in the Riemann Normal Coordinates set up around the ligthcone vertex $p$. We compute the expansions up to the leading curvature correction and the higher order terms are irrelevant for the small sphere limit in non-vacuum. Firstly, we need to fix a gauge for the null generators $\ell^\pm$.

We choose the leading contribution $\l^\pm_\mu$ to the outer and inner null generators as
\beq
\l^{+\mu} := (1, n^i),\;\; \l^{-\mu} :=\frac12(1,-n^i),
\eeq
the RNC expansions of $\ell^\pm_\mu$ restricted on $S_l$ are then given by:
\beq
\begin{aligned}
\ell^{+\mu} =& (1, n^i) = \l^{+\mu}, &\ell^+_\mu =& (-1, n^i) = \l^+_\mu , \\
\ell^{-\mu} =& \l^{-\mu}+\frac{l^2}{6}R_{+-+-}\ell^{+\mu}+\frac{l^2}{3}R\indices{_{+-+}^\mu}+O(l^3),&\ell^{-}_\mu =& \l_{\mu}^-+\frac{l^2}{6}R_{+-+-}\ell^+_{\mu}+O(l^3),
\end{aligned}
\eeq
where $n^i$ is a normalized spacelike vector indicating the spatial direction, $\ell^\pm_{\mu}=g_{\mu\nu}\ell^{\pm\nu}, \l^\pm_{\mu}=\eta_{\mu\nu}\l^{\pm\nu},$ and we use abbreviations such as $R_{+-+-}=R(\l^+,\l^-,\l^+,\l^-),$ etc.

We can then compute the expansions directions from the definitions: 
\begin{enumerate}

\item The expansions on $S_l$ are
\begin{align}\label{expansions}
\theta^+(l) =& \frac{n-2}{l} -\frac{l}{3}R_{+-} +O(l^3),\\
\theta^-(l)=&-\frac{n-2}{2l}-\left(\frac23 R_{+-}+\frac16 R_{++}-\frac{n+2}{6}R_{+-+-}\right)l+ O(l^3).
\end{align} 

\item The twists on $S_l$ are
\beq\label{twists}
\begin{aligned}
\omega^+_\mu(l) =& \frac{l}3 R_{\mu +-+}+O(l^3),\\
\omega^-_\mu(l) =& -\omega^+_\mu. 
\end{aligned} 
\eeq

\item The Ricci scalar on $S_l$ is
\beq\label{ricci}
\begin{aligned}
\R(l) =& \frac{(n-2)(n-3)}{l^2}+R+\frac{4n}{3}R_{+-}-\frac{n(n-1)}{3}R_{+-+-}+O(l^2).
\end{aligned}
\eeq
\end{enumerate}
To apply the algorithm on the light cone cuts, we choose $\sigma$ to be our light cone cuts $S_l$ and we are interested in the limit $l\rightarrow 0$. We only discuss in detail the case for $n\geq 6$, and leave out the detailed calculations for $n=3,4,5$ which are very similar to the general case. We only do the non-vacuum case here and leave the vacuum case to future works. 

In general, it is complicated to exactly solve this polynomial equation and we are not guaranteed to have a closed-form solution. Since we only work in the perturbative regime, a solution in expansion form suffices. Even though a solution $\xi_0$ with the leading curvature perturbation is enough for the non-vacuum limit, it is still tricky to compute the integral with $\xi_0^{2-n}$, as we will see that $\xi_0^{2-n}$ contains curvature terms raised to non-integer powers. Furthermore, we might need to choose a particular gauge by hand in order to satisfy the condition 1 (\ref{condition1}) above.  Hence, instead of treating the general case, we choose to only evaluate the small sphere limits for those lightcone cuts $\{(p,e_0)\}$ which enjoy the spherical symmetry approximately up to the leading order of curvature correction\footnote{Note that our case is more general than the round spheres in spherically symmetric spacetimes studied by NR, so we cannot directly apply the simplified constraint equations developed in~\cite{nomura2018area}. }. The advantage is that we can now take a short-cut by spherically averaging each coefficient in the constraint equation (\ref{eq:polynomial}) before solving for $\xi_0$. 
It turns out that according to our gauge choice is a good one, as all the coefficients $\epsilon_i$ vanish up to the leading order of curvature correction. 
\begin{equation}
\overline{\epsilon_i} := \frac{1}{\Omega_{n-2}}\int_{S^{n-2}} \epsilon_i \;\dd \Omega_{n-2} = 0 + O(l^2),\; \text{for}\; i=1,2,3,4,5.
\end{equation}
To show these, we first compute the relevant quantities listed in (\ref{eq:coeffi}) using our data (\ref{expansions},\ref{twists},\ref{ricci}). 
\beq
\begin{aligned}
\D\log\theta^+ =& \frac{-l^2}{3(n-2)}\D R_{++} = \frac{-2l}{3(n-2)} (R_{\mu+}+R_{++}\ell^-_\mu +R_{+-}\ell^+_\mu)+O(l^2),\\
\Box\log\theta^+ =& \frac{-2}{3(n-2)}(R-\frac{n-2}{2}R_{++}+nR_{+-})+O(l^2),  \\
\D\cdot\omega =& \frac13\left((n-1)R_{-+-+} -R_{+-} - \frac12 R_{++}\right)+O(l^2) , \\ 
\omega\cdot\D\log\theta^+ =&\, O(l^2).
\end{aligned}
\eeq

These following terms vanish at the leading order,
\beq\label{averagingcoeffi2}
\begin{aligned}
\overline{\epsilon_{1}} & =\frac{-2}{3(n-4)\Omega_{n-2}}\int_{S^{n-2}}(R-\frac{n-2}{2}R_{++}+nR_{+-})\,\dd\Omega_{n-2}+O(l^2)=O(l^2), \\
\overline{\epsilon_{3}} & =-\frac{(n-2)}{\Omega_{n-2}}\int_{S^{n-2}} (\D\cdot\omega-\Box\log\theta^+ -(n-2)\omega\cdot\D\log\theta^+ +(n-2)| \D\log\theta^+ |^2))\dd\Omega_{n-2},\\
& = -\frac{(n-2)}{3\Omega_{n-2}}\int_{S^{n-2}}(n-1)R_{-+-+} -R_{+-} - \frac12 R_{++} \dd\Omega_{n-2} +O(l^2) = O(l^2), \\
\overline{\epsilon_{2}} & = O(l^{2}), \;\;\;\;\overline{\epsilon_{4}} = O(l^2), \;\;\;\overline{\epsilon_{5}}  = O(l^2).
\end{aligned}
\eeq

The non-vanishing terms are
\beq \label{averagingcoeffi}
\begin{aligned}
\overline{\theta^+\theta^-} =& \frac{1}{\Omega_{n-2}}\int_{S^{n-2}}  -\frac{(n-2)^2}{2l^2} -\frac{n-2}{6}\left(4R_{+-}-(n+2)R_{+-+-}\right)  \dd\Omega_{n-2} + O(l^2) ,\\
=& -\frac{(n-2)^2}{2l^2} -\frac{(n-2)[(n-6)Ric(e_0,e_0)-2R]}{6(n-1)} + O(l^2).\\
\\
\overline{\rho} =& \frac{-(n-2)}{2(n-3)\Omega_{n-2}} \int_{S^{n-2}}  \frac{(n-2)(n-3)}{l^2}+R+\frac{4n}{3}R_{+-}-\frac{n(n-1)}{3}R_{+-+-}\;  \dd\Omega_{n-2} + O(l^2),\\
=& -\frac{(n-2)^2}{2l^2}-\frac{(n-2)[nRic(e_0,e_0)+R]}{6(n-1)}+ O(l^2).
\end{aligned}
\eeq
It yields a simple form of the constraint equation, and it turns out this simplified constraint equation is identical as the one for round spheres in spherical symmetric spacetime~\cite{nomura2018area}. We are left with
\begin{equation}\label{constraintsimple}
(\overline{\theta^+\theta^-}-\overline{\rho})\xi^{n-1}+\overline{\rho}\xi^2=0.
\end{equation}
We can easily solve (\ref{constraintsimple}),
\beq
\begin{aligned}\label{eq:solutionxi}
\xi_0^{3-n} &= 1 - \frac{\overline{\theta^+\theta^-}}{\overline{\rho}} = 2\frac{n-3}{n-2}\frac{\overline{\theta^+\theta^-}}{\overline{\R}} +1
=\frac{2l^2G(e_0,e_0)}{(n-2)(n-1)}
=\frac{2l^2\Omega_{n-2}G_N T(e_0,e_0)}{n-1} +O(l^3)
\end{aligned}
\eeq

Condition 1 \ref{condition1} is trivially satisfied because of the spherical symmetry so we only need to check condition 2 \ref{condition2}. The cross-focusing equation (\ref{eq:constraintk}) applied on $Y_0$ gives
\beq
\partial_-\theta^+[Y_0]=\partial_+\theta^-[Y_0] - 2\D\cdot\omega^{+}[Y_0].
\eeq
In BNR \cite{bousso2019outer}, it is shown that 
\beq
\begin{aligned}
\partial_+\theta^-[Y_0] 
  &= -\frac{1}{2}\xi_0^{2}\R + \left[\xi_0^{2n-2}-n\xi_0^{n+1}+(n-2)\xi_0^{n}+2\xi_0^{4}-\xi_0^{3}\right] |\D\log\theta^+ |^{2} \\
  &\qquad -\left[2\xi_0^{2n-2}-n\xi_0^{n+1}+(n-2)\xi_0^{n}\right]\omega^+\cdot\D\log\theta^+ \\
  &\qquad + \left(\xi_0^{3}-\xi_0^n\right)\Box\log\theta^+
    + \xi_0^{n}\D\cdot\omega^+ + \xi_0^{2n-2}\omega^{+2},\\  
\end{aligned}
\label{eq:qeq}
\eeq
and
\beq
\begin{aligned}
  -2\D\cdot\omega[Y_0] &= 2\left(\xi_0-\xi_0^2\right)\left[\xi_0^2 - (n-2)\xi_0^{n-1}\right] |\D \log\theta^+ |^2 
\\&\qquad  - 2(n-2)\left(\xi_0^{n+1}-\xi_0^n\right) \omega^{+}\cdot\D \log\theta^+ \\
 &\qquad - 2\left(\xi_0^3-\xi_0^n\right) \Box\log\theta^+ - 2\xi_0^n {\cal D}\cdot\omega^+.
\end{aligned}
\eeq
In our case, we should substitute in the averaged data that we just calculated (\ref{averagingcoeffi2},\ref{averagingcoeffi}). It is then again a matter of power counting for $n\geq 6$:
\beq
\partial_-\theta^+[Y_0]=\partial_+\theta^-[Y_0] - 2\D\cdot\omega^{+}[Y_0] = -\frac{1}{2}\xi_0^{2}\R[S_l] + O(l^{>\frac{2n-2}{3-n}}).
\eeq
where the leading term has order $O(l^\frac{2n-2}{3-n})$ and it is less than zero for sufficiently small light cone cuts $S_l$. Hence, condition 2 is also satisfied.

According to (\ref{hrtarea}), the outer entropy is therefore 
\beq
\S = \frac{1}{4G_N\hbar}\int_{S_l}\xi_0^{n-2}\,\dd \sigma= \frac{A(S_l)}{4G_N\hbar \xi_0^{n-2}}
\eeq
where the area $A(S_l)$ is given by
\beq
A(S_l) = \Omega_{n-2}l^{n-2}.
\eeq
Hence, we have
\beq
\S=\frac{\Omega_{n-2}l^{n-2}}{4G_N\hbar}\left(\frac{2l^2\Omega_{n-2}G_N T(e_0,e_0)}{n-1}\right)^{\frac{n-2}{n-3}}.
\eeq

This concludes our small sphere limit calculation. Note that if we assume the validity of the HRRT prescription, we no longer need the holographic duality in order to define and evalute the outer entropy, and we can work in a spacetime that is aymptotically flat. It would be interesting to ask what the outer entropy tends to as $\sigma$ apporaches the spatial infinity in an aymptotically flat manifold, and if it has any relation with the ADM mass. 

\bibliographystyle{JHEP}
\bibliography{outer}

\end{document}